\documentclass[manuscript,screen,nonacm]{acmart}
\AtBeginDocument{%
  }
\usepackage{adjustbox}

\begin{document}

\title{\textbf{Feature Anchors for Time-Series Sensor-Based Human Activity Recognition}}
\author{Ruijie Yao}
\affiliation{%
  \institution{Mechanical Engineering \& Materials Science, Duke University}
  \city{Durham}
  \state{North Carolina}
  \country{United States}}
\email{ruijie.yao@duke.edu}

\author{Chenhang Li}
\affiliation{%
  \institution{Mechanical Engineering \& Materials Science, Duke University}
  \city{Durham}
  \state{North Carolina}
  \country{United States}}

\author{Danyang Zhuo}
\affiliation{%
  \institution{Computer Science, Duke University}
  \city{Durham}
  \state{North Carolina}
  \country{United States}}
\email{danyang@cs.duke.edu}

\author{Tingjun Chen}
\affiliation{%
  \institution{Electrical and Computer Engineering, Duke University}
  \city{Durham}
  \state{North Carolina}
  \country{United States}}
\email{tingjun.chen@duke.edu}

\author{Xiaoyue Ni}
\affiliation{%
  \institution{Mechanical Engineering \& Materials Science, Duke University}
  \city{Durham}
  \state{North Carolina}
  \country{United States}}
\email{xiaoyue.ni@duke.edu}
\renewcommand{\shortauthors}{R. Yao et al.}


\begin{abstract}
Wearable Human Activity Recognition (HAR) still lacks a representation that is both explicit and adaptable. Handcrafted time-series features (TSFs) capture meaningful motion statistics and remain competitive on standard benchmarks, but they are usually used as fixed preprocessing outputs. Deep models learn adaptable representations directly from raw signals, but those representations are typically latent and difficult to inspect.
We address this gap by treating handcrafted TSFs as \textit{feature anchors}: explicit intermediate representations that remain inside the model and are adjusted by neural context instead of being discarded. We propose the Temporal Conditioning Network for Feature Anchors (TCNet), which extracts handcrafted anchors, encodes complementary time-domain and frequency-domain context from raw IMU windows, and predicts context-conditioned scale, bias, and gating parameters to modulate anchor groups directly in feature space. This design keeps anchor semantics visible while allowing the representation to adapt to the classification objective.
Across five HAR benchmarks, TCNet achieves 70.2\% mF1 on USC-HAD, 85.1\% mF1 on Daphnet, 93.9\% mF1 on MHealth, and 94.5\% mF1 on PAMAP2. Relative to rTsfNet, it improves by 4.5 points on USC-HAD, 14.6 points on Daphnet, and 6.5 points on MHealth. Ablations show that the gains come primarily from anchor guidance rather than simple branch fusion, and feature-space analyses indicate that several discriminative TSF families are not reliably accessible in standard latent representations. These results suggest that, for HAR, handcrafted TSFs are most useful when they remain explicit and adaptable within the model.
The code is available at: https://github.com/ni-x-lab/TCNet-har
\end{abstract}



\maketitle

\section{Introduction}
\label{sec:intro}

\begin{figure*}[t]
  \centering
  \includegraphics[width=\textwidth]{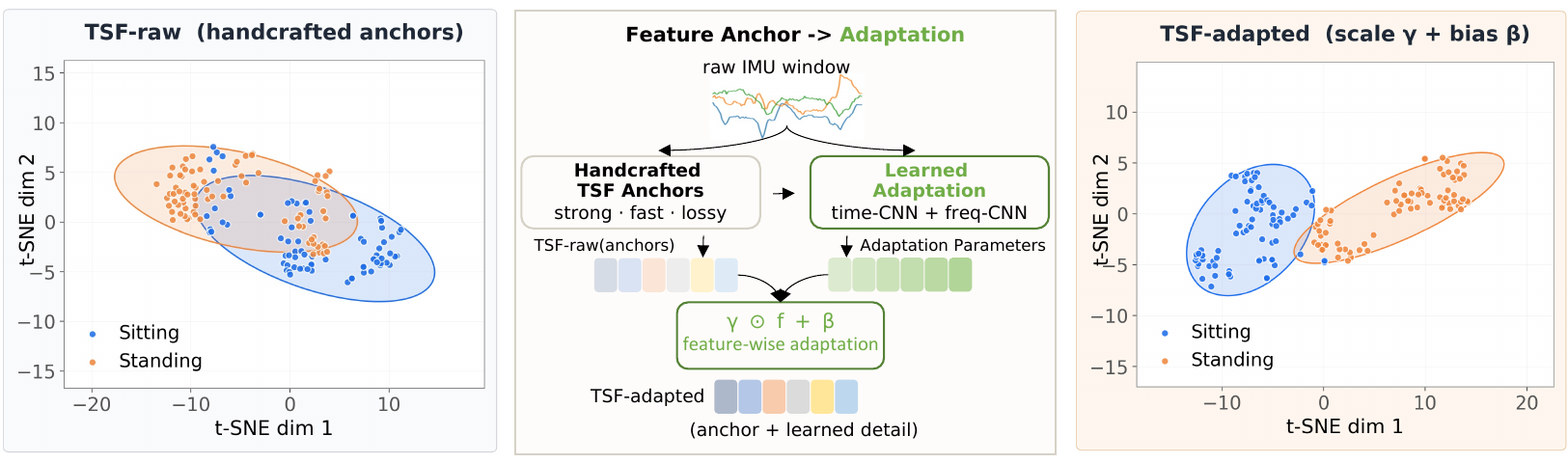}
\caption{
From handcrafted feature anchors to adapted representations.
Handcrafted time-series features (TSFs) are treated as \emph{feature anchors} that provide an explicit statistical scaffold but are not fully discriminative (left).
Instead of replacing them, TCNet learns a lightweight \emph{adaptation} (feature-wise scale $\gamma$ and bias $\beta$) conditioned on raw sensor context to refine these anchors (middle).
This results in adapted anchors that improve class separability while remaining interpretable (right).
The plots show t-SNE embeddings of the corresponding features on the UCI-HAR test set, where relative positions reflect neighborhood structure rather than absolute coordinates.
}
  \label{fig:anchor_analogy}
  \Description{A teaser figure illustrating handcrafted time-series features as feature anchors, their context-conditioned adaptation with scale and bias in TCNet, and the resulting improvement in class separability on the UCI-HAR test set.}
\end{figure*}

Wearable Human Activity Recognition (HAR) infers activities such as walking, climbing, and freezing episodes from inertial sensor streams~\cite{zhang2012usc,reiss2012introducing,anguita2013public,bachlin2009wearable,bulling2014tutorial,lara2013survey}. It supports applications such as fall detection~\cite{igual2013fall}, Parkinson's disease monitoring~\cite{bachlin2010wearable}, fitness tracking~\cite{shoaib2015survey}, and context-aware assistance~\cite{haresamudram2025survey}. In these settings, accuracy must coexist with memory, latency, and compute constraints, since wearable HAR models are often expected to run on edge devices.

\begin{figure*}[t]
  \centering
  \includegraphics[width=\linewidth]{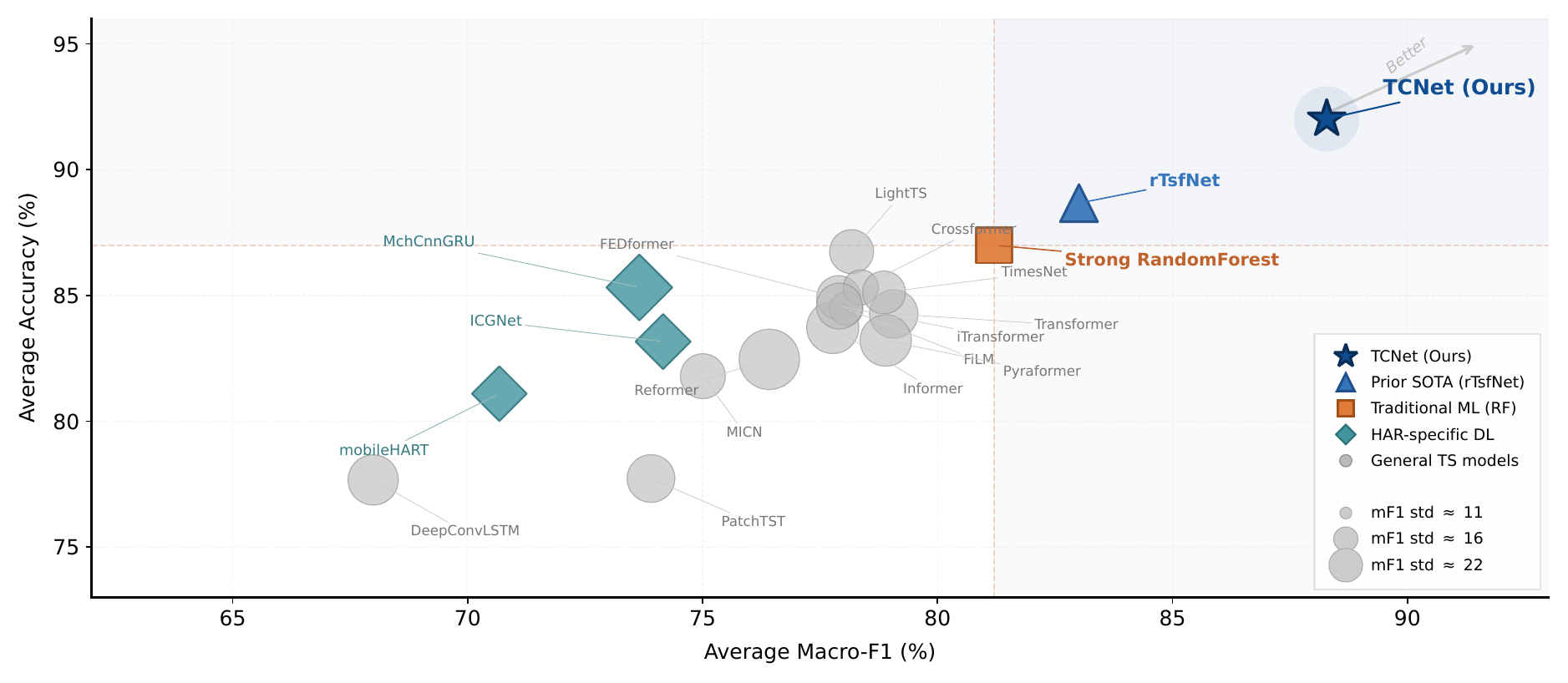}
  \Description{Bubble chart comparing HAR methods on five benchmarks. The horizontal axis is mean accuracy across datasets; the vertical axis is mean macro-F1; bubble area encodes how many datasets each method ranks best or second-best. TCNet appears in the upper-right, above deep baselines and the RF-TSF reference.}
  \caption{Overall performance comparison across five HAR benchmarks. Each bubble represents one method; horizontal position indicates mean accuracy, vertical position indicates mean macro-F1, and bubble area reflects the number of datasets where the method ranks first or second. TCNet (top-right) surpasses both deep baselines and the RF-TSF reference trained on the same handcrafted features (Sec.~\ref{sec:rf_baseline}), indicating that context-conditioned anchor adaptation provides gains beyond those of fixed handcrafted pipelines.}
  \label{fig:tcnet_bubble}
\end{figure*}

Handcrafted time-series features (TSFs) remain a strong starting point. As Fig.~\ref{fig:anchor_analogy} illustrates, they provide an explicit statistical scaffold by organizing motion signals into semantically meaningful feature groups. Features such as spectral concentration, crossing behavior, and autocorrelation summarize motion statistics in a compact and interpretable form~\cite{preece2009comparison,christ2018tsfresh}. More importantly, they remain highly competitive. In our experiments, a strengthened Random Forest trained on the same two-scale TSF representation used by our Temporal Conditioning Network for Feature Anchors (TCNet) reaches 92.6\% mF1 on UCI-HAR and 89.4\% mF1 on PAMAP2, outperforming many end-to-end deep baselines while using a much smaller model. Figure~\ref{fig:tcnet_bubble} places this result in context: handcrafted-feature pipelines remain competitive, and TCNet improves further by adapting those same features rather than discarding them. This suggests that handcrafted TSFs still capture much of the structure needed for HAR and remain attractive for edge deployment.

Deep HAR models, in contrast, learn representations directly from raw signals, from CNN/RNN encoders~\cite{plotz2011feature,hammerla2016deep} to HAR-specific architectures such as ICGNet~\cite{dua2023inception}, MchCnnGRU~\cite{lu2022multichannel}, and Transformer variants such as mobileHART~\cite{ek2023transformer}. Recent systems also rely on heavier backbones to push performance further, including SensorLLM~\cite{sensorllm2025} and rTsfNet~\cite{rTsfNet2024}. These advances confirm the value of learned representations, but they also raise a practical question for wearable HAR: if handcrafted TSFs are already strong and compact, do we need to relearn the entire representation from raw signals with increasingly large networks?

We argue that the answer is often no. The limitation of handcrafted TSFs is not that they are obsolete, but that they are \emph{lossy}. Window-level extraction preserves useful motion summaries while discarding fine-grained, task-relevant detail. If TSFs already provide most of the statistical scaffold needed for recognition, then the neural model need not rebuild a full representation from scratch. A lightweight model only needs to infer the missing detail and add it back in a task-aware way. Figure~\ref{fig:anchor_analogy} visualizes this idea: TCNet starts from handcrafted feature anchors, predicts a lightweight feature-wise adaptation from raw-signal context, and produces adapted anchors that are more separable while remaining interpretable.

This leads to our central formulation: \textbf{handcrafted TSFs as feature anchors}. Rather than treating TSFs as fixed preprocessing outputs or auxiliary side information, we keep them inside the model as the main intermediate representation. A lightweight network reads the raw IMU window, extracts time-domain and frequency-domain context, and predicts feature-space modulation parameters that adapt the anchors to the current input. The model therefore does not discard handcrafted features; it refines them.

TCNet instantiates this idea through context-conditioned anchor adaptation. It first computes handcrafted TSFs as explicit anchors, then uses lightweight time-domain and frequency-domain branches to predict scale, bias, and gating parameters that reshape anchor groups directly in feature space. This preserves the efficiency and interpretability of handcrafted representations, since the anchors remain visible throughout the forward pass, while reducing the burden on the neural model, which adapts a strong prior instead of relearning the full representation from raw data.

Experiments on five HAR benchmarks show that this strategy is effective. As summarized in Fig.~\ref{fig:tcnet_bubble}, TCNet occupies the top-right region of the accuracy--macro-F1 tradeoff, surpassing both deep baselines and the strong RF-TSF reference built on the same handcrafted features. TCNet achieves the best macro-F1 on USC-HAD~\cite{zhang2012usc}, Daphnet~\cite{bachlin2009wearable}, MHealth~\cite{banos2014mhealthdroid}, and PAMAP2~\cite{reiss2012introducing}, improving over rTsfNet by 4.5 points on USC-HAD, 14.6 points on Daphnet, and 6.5 points on MHealth. Ablations show that these gains come primarily from adapting handcrafted anchors rather than from simple branch fusion, while feature-space analyses show that the adapted anchors remain recognizable and more discriminative. Together, these results suggest that strong handcrafted features need not be replaced by larger networks; they can instead serve as compact anchors that a lightweight model refines efficiently.
Contributions are as follows:
\begin{itemize}
    \item \textbf{Feature anchors for efficient HAR:} We formulate handcrafted TSFs as strong intermediate representations for wearable HAR and argue that they should be adapted rather than discarded.
    \item \textbf{Context-conditioned anchor adaptation:} We introduce TCNet, a lightweight model that predicts scale, bias, and gating parameters from raw IMU context to refine handcrafted feature groups directly in feature space.
    \item \textbf{Empirical evidence for anchor-based learning:} Across five HAR benchmarks, we show that TCNet delivers strong accuracy with a substantially smaller model than heavy baselines, and that its gains arise from anchor adaptation rather than simple multi-branch fusion.
\end{itemize}
\section{Related Work}
\label{sec:related}

\subsection{Handcrafted Features as Strong and Efficient HAR Representations}
Early wearable HAR systems commonly relied on handcrafted time-series features (TSFs) together with shallow classifiers such as Random Forests, SVMs, and decision trees~\cite{bulling2014tutorial, haresamudram2025survey}. Feature extraction libraries such as \textit{tsfresh}~\cite{christ2018tsfresh} further standardized this workflow by exposing a large catalog of explicit window-level descriptors. Although later studies showed that convolutional and recurrent models can match or exceed classical pipelines~\cite{hammerla2016deep, guan2017ensembles}, handcrafted TSFs have remained attractive for two reasons. First, they provide compact and interpretable summaries of motion statistics, including spectral concentration, crossing behavior, and autocorrelation structure~\cite{preece2009comparison, christ2018tsfresh}. Second, they support lightweight deployment, which remains important in wearable and edge settings.

\begin{figure}[t]
  \centering
  \includegraphics[width=\linewidth]{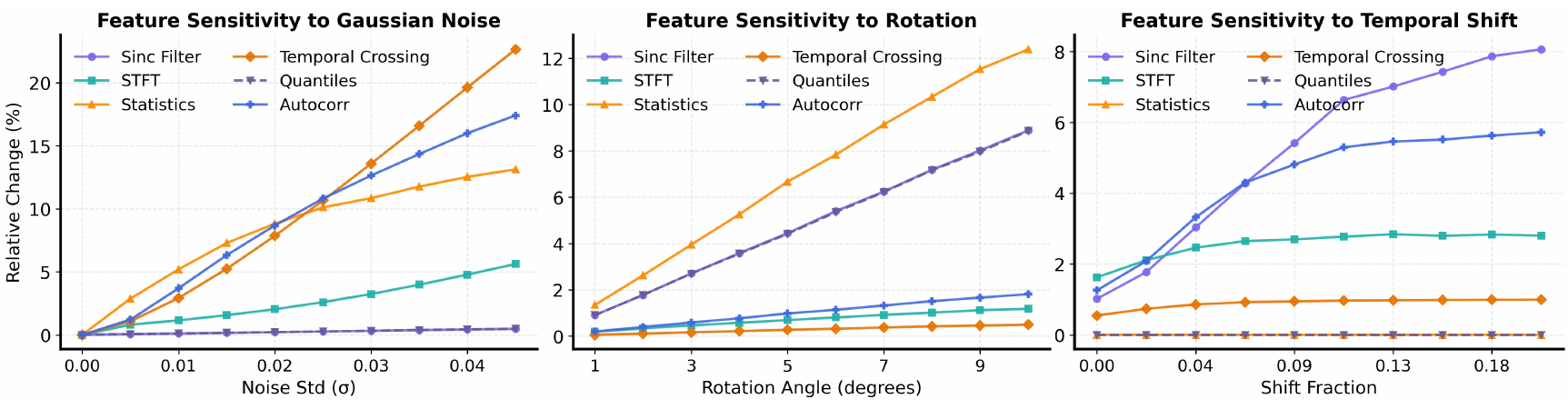}
\caption{
Handcrafted TSFs are highly sensitive to small input perturbations. Even mild noise ($\sigma \leq 0.04$), rotation ($< 10^\circ$), or temporal shift ($< 0.2$) can cause feature changes exceeding 20\%, 12\%, and 8\%, respectively. The responses also vary across TSF families, indicating that handcrafted features are not consistently robust. This instability limits their effectiveness as fixed representations and motivates adaptive modeling.
}
  \label{fig:feature_sensitivity}
\end{figure}

At the same time, strong handcrafted features are not sufficient as fixed representations. Figure~\ref{fig:feature_sensitivity} shows that different TSF families respond quite differently to common perturbations such as Gaussian noise, rotation, and temporal shift. Some families remain relatively stable, whereas others change sharply under specific perturbation regimes. This means that the usefulness of a handcrafted feature is context-dependent: a TSF family that is discriminative in one setting may become unreliable in another. As a result, directly using TSFs as fixed preprocessing outputs can leave performance on the table, because the representation cannot correct for input-dependent distortion or recover detail lost during feature summarization.
Our work builds on the strength of handcrafted TSFs but departs from the standard classical pipeline. Rather than treating TSFs as fixed preprocessing outputs for a shallow classifier, we treat them as strong intermediate representations that should remain inside the model and be adapted when the input context requires it.

\subsection{Deep HAR from Raw Signals}
Deep HAR models learn representations directly from raw sensor streams, including CNNs, recurrent architectures, and Transformer-based models~\cite{deepconvlstm2016, vaswani2017attention, wang2019deep}. Early frameworks such as DeepSense~\cite{yao2017deepsense} showed that end-to-end deep pipelines can replace manual preprocessing across a range of mobile sensing tasks. Subsequent work improved discriminability through multi-level attention~\cite{ma2019attnsense}, adversarial training~\cite{abedin2021attend}, and structured augmentation for convolutional encoders~\cite{shao2023convboost}. Recent works have further advanced multi-sensor fusion and sequence modeling with architectures, such as IF-ConvTransformer~\cite{zhang2022ifconvtransformer}, DynamicWHAR~\cite{miao2022dynamicwhar}, multi-modal temporal segment attention~\cite{gao2023mmtsa}, masked autoencoders for multi-device sensing~\cite{miao2023stmae}, and cross-modal recognition with IMUZero~\cite{imuzero2025}.

These methods demonstrate the strength of learned representations from raw signals, but they typically place the full burden of representation learning on the network itself. Recent systems also increasingly rely on heavier backbones to push performance further, including LLM-based or large-capacity variants such as SensorLLM~\cite{sensorllm2025}. Our work asks a different question: if handcrafted TSFs already capture much of the useful structure for HAR, can a lightweight model adapt them instead of relearning the full representation from scratch?

\subsection{Hybrid HAR Methods and the Missing Adaptation Formulation}
Several HAR systems have combined handcrafted and neural components, showing that handcrafted priors remain useful even in deep pipelines. Prior hybrid designs include multi-task objectives that combine handcrafted and neural branches~\cite{peng2018aroma}, signal rotation alignment before feature extraction as in rTsfNet~\cite{rTsfNet2024}, and temporal localization priors that guide feature extraction~\cite{bock2024temporal}. These methods demonstrate that handcrafted structure can still improve performance, but they typically use TSFs as preprocessing outputs, side information, or auxiliary branches whose role is fixed before classification.

This leaves a specific representation gap. Existing hybrid HAR methods rarely keep handcrafted TSFs as the \emph{main explicit intermediate representation} and then adapt them directly with neural context during learning. Our work targets exactly this missing formulation: instead of appending handcrafted features to a learned representation, TCNet keeps handcrafted feature groups visible throughout the forward pass and refines them directly in feature space.

\subsection{Structured Priors, Feature Modulation, and Explicit Intermediate Representations}
Our formulation is also related to two broader ideas outside standard HAR pipelines. First, structured priors can simplify high-dimensional learning by constraining the search space. In object detection, anchor boxes encode geometric priors that stabilize optimization and reduce the burden of unconstrained prediction~\cite{ren2015faster, girshick2014rcnn}. Second, many neural architectures adapt internal representations by predicting feature-wise affine transformations from context, including Batch Normalization~\cite{ioffe2015batch}, Layer Normalization~\cite{ba2016layer}, FiLM~\cite{perez2018film}, and conditional normalization~\cite{dumoulin2017learned}.

TCNet combines these ideas in a different setting. Instead of applying context-conditioned modulation only to latent channels, it applies it to handcrafted feature groups that remain explicit throughout the forward pass. In this sense, our method is not simply another hybrid HAR model; it is a structured-prior formulation in which handcrafted TSFs serve as feature anchors and are refined, rather than discarded, by a lightweight neural module.

\section{Methodology: Context-Conditioned Feature Anchor Learning}
\label{sec:method}

TCNet is built around a simple representation principle: handcrafted time-series features should remain explicit inside the network, but they should not remain fixed. Instead of asking a deep backbone to rediscover these motion statistics from raw signals, TCNet treats them as \emph{feature anchors} and learns a lightweight, context-conditioned correction directly in feature space.

Given an IMU window \(x \in \mathbb{R}^{B \times C \times L}\), TCNet operates in four stages. First, it partitions the input into non-overlapping temporal blocks at one or more block sizes. Second, it extracts global time-domain and frequency-domain context from the raw window. Third, for each block, it computes a differentiable anchor tensor composed of multiple TSF families. Fourth, it predicts bounded affine corrections and interpolation gates for these anchors, produces multiple corrected anchor views, and aggregates them across views, sensor groups, blocks, and scales for classification.

\subsection{Input and Multi-Scale Blocking}

Let \(x \in \mathbb{R}^{B \times C \times L}\) denote a batch of multichannel IMU windows, where \(B\) is the batch size, \(C\) is the number of sensor channels, and \(L\) is the window length. For each configured block size \(m\), we unfold the temporal dimension into
\[
N_m = \left\lfloor \frac{L-m}{s_m} \right\rfloor + 1
\]
non-overlapping or strided blocks, where \(s_m\) is the stride for that scale. Each block is processed independently by the differentiable TSF extractor, and each scale produces its own block-level representation. Using multiple block sizes allows the model to capture both short transitions and longer activity patterns within the same input window.

\begin{figure}[t]
  \centering
  \includegraphics[width=\linewidth]{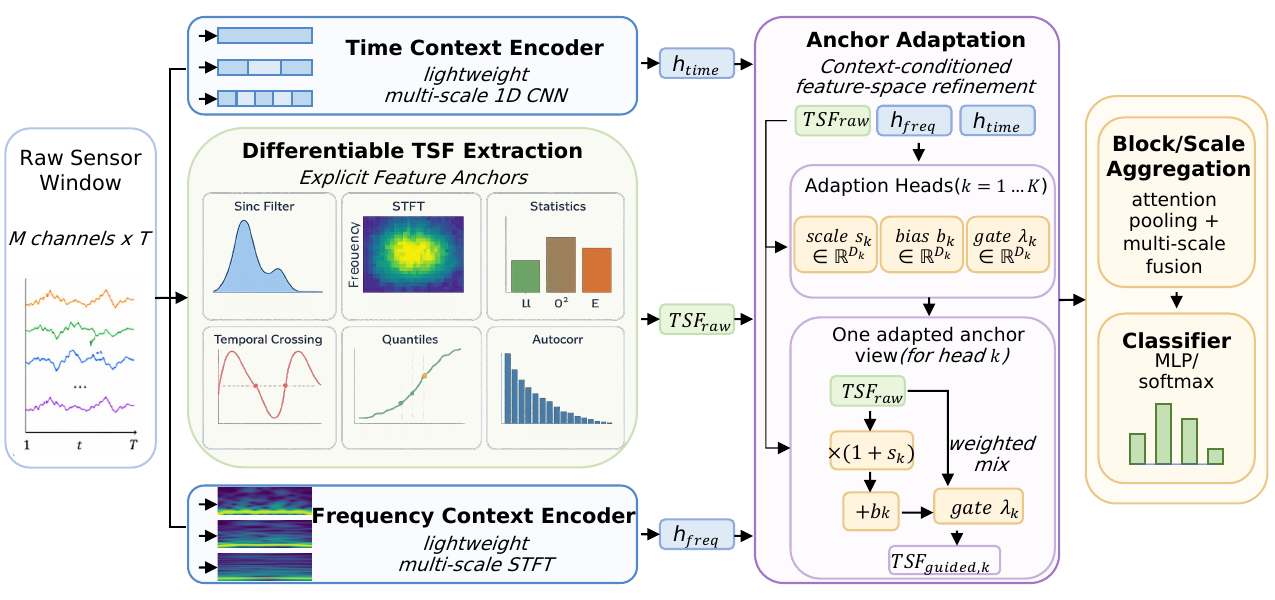}
\caption{
TCNet architecture. From a raw IMU window, TCNet computes three representations in parallel: (i) a lightweight time-context encoder producing $h_{time}$, (ii) a differentiable TSF extractor producing the raw anchor tensor $TSF_{raw}$, and (iii) a lightweight frequency-context encoder producing $h_{freq}$. The concatenated context embeddings condition an anchor-adaptation module, which predicts per-group scale $s$, bias $b$, and gating $\lambda$ to generate $K$ adapted anchor views from $TSF_{raw}$ directly in feature space. These adapted anchors are then aggregated across temporal blocks and window scales before classification. Unlike standard latent-feature pipelines, the handcrafted anchors remain explicit throughout the forward pass, so prediction is based on a statistically grounded representation that is refined, rather than replaced, by neural context.
}
  \label{fig:framework}
\end{figure}

\subsection{Global Context from Raw Signals}

The correction applied to an anchor should depend on the raw signal context of the current window. TCNet therefore uses two lightweight branches that operate directly on \(x\): a temporal branch and a frequency branch.

\subsubsection{Temporal Context Branch}

The temporal branch is a lightweight 1D convolutional encoder. It applies one or more temporal convolutions over the raw multichannel signal and then performs global average pooling to obtain a compact time-domain embedding:
\[
h_{\mathrm{time}} = f_{\mathrm{time}}(x) \in \mathbb{R}^{B \times d_t}.
\]
This branch captures local motion transitions and waveform patterns that may not be preserved after handcrafted summarization.

\subsubsection{Frequency Context Branch}

The frequency branch first converts the raw signal into a log-magnitude spectrogram using STFT:
\[
S = \log(1 + |\mathrm{STFT}(x)|^2).
\]
A \(1\times1\) convolution mixes channels in the spectrogram domain, after which a lightweight MLP-Mixer aggregates patch-level spectral information. Global averaging then produces a compact frequency-domain embedding:
\[
h_{\mathrm{freq}} = f_{\mathrm{freq}}(x) \in \mathbb{R}^{B \times d_f}.
\]
This branch captures spectral regularities such as periodicity, concentration, and harmonic structure that are useful for activity discrimination.

The final global context is the concatenation
\[
h_{\mathrm{global}} = [h_{\mathrm{time}}; h_{\mathrm{freq}}] \in \mathbb{R}^{B \times (d_t+d_f)}.
\]

\subsection{Differentiable Feature Anchors}

For each temporal block, TCNet computes a differentiable handcrafted feature representation rather than a latent feature map. Specifically, for a block tensor \(x^{(m)} \in \mathbb{R}^{B \times N_m \times m \times C}\), the differentiable TSF extractor produces
\[
Z^{(m)}_{\mathrm{raw}} \in \mathbb{R}^{B \times N_m \times C \times D},
\]
where \(D\) is the total anchor dimensionality.

Unlike a minimal handcrafted pipeline, the differentiable extractor in TCNet is deliberately designed as a \emph{multi-family anchor generator}. Rather than relying on a small set of coarse statistics, it combines several complementary feature families, each capturing a different aspect of motion dynamics while remaining differentiable and thus trainable end-to-end:

\begin{itemize}
    \item \textbf{Filterbank features.}
    We first include a differentiable Sinc-based filter bank, which acts as a learnable band-pass decomposition of the raw sensor signal. Each filter extracts energy from a specific frequency band, producing a compact summary of how motion intensity is distributed across frequency ranges. Compared with a fixed FFT summary, this component provides a more flexible spectral scaffold while preserving the interpretation of band-limited responses.

    \item \textbf{Spectral features.}
    We further compute spectral anchors using a differentiable STFT with a learnable Gaussian window. From this representation, the extractor derives frame-aggregated magnitude features as well as phase-related summaries. These features capture periodicity, spectral concentration, and oscillatory structure, which are particularly informative for distinguishing repetitive activities such as walking, jogging, or cycling from less periodic motions.

    \item \textbf{Basic statistical features.}
    To preserve coarse amplitude-level information, the extractor includes standard descriptive statistics such as mean, soft minimum, soft maximum, root-mean-square (RMS), and standard deviation. These statistics summarize signal location, range, energy, and dispersion, which remain strong cues for separating static from dynamic activities and for characterizing overall motion intensity.

    \item \textbf{Higher-order distributional shape features.}
    Beyond first- and second-order moments, the extractor also computes adaptive skewness and kurtosis features. These features describe the asymmetry and peakedness of the signal distribution and are implemented through differentiable moment estimation. They provide a richer description of waveform shape and can capture subtle differences in activity-specific signal morphology that are missed by mean-variance summaries alone.

    \item \textbf{Temporal crossing features.}
    To explicitly encode temporal switching behavior, TCNet includes a family of differentiable crossing-based features. These consist of soft zero-crossing rate, mean-crossing rate, zero-crossing rate of the first-order difference, a temporal regularity statistic, and a soft extrema count. Together, these features characterize how frequently the signal changes sign, how often it crosses its mean level, how rapidly local slope patterns switch, and how oscillatory or jagged the motion is over time.

    \item \textbf{Quantile features.}
    The extractor also includes differentiable quantiles computed through soft ranking. In contrast to mean-based summaries, quantile features are more robust to outliers and provide a compact description of the signal distribution at multiple percentile levels. They are useful for representing amplitude spread and tail behavior while remaining stable under moderate noise.

    \item \textbf{Autocorrelation features.}
    Finally, we compute differentiable autocorrelation coefficients at multiple lags. These features measure temporal self-similarity and periodic structure, allowing the anchor tensor to preserve rhythm and repetition patterns directly in feature space. Such lag-based dependencies are particularly important for cyclic activities whose discriminative structure lies not only in amplitude or spectrum, but also in repeating temporal organization.
\end{itemize}

Taken together, these differentiable TSF families produce a structured anchor tensor that retains complementary motion statistics in explicit form. Rather than collapsing these statistics into an opaque latent embedding, TCNet preserves them as semantically organized feature anchors and later refines them through context-conditioned correction.s

\subsection{Context-Conditioned Anchor Correction}

The core of TCNet is a block-aware correction module that refines raw anchors using raw-signal context. For each scale \(m\), the module takes the raw anchor tensor \(Z^{(m)}_{\mathrm{raw}}\) together with \(h_{\mathrm{global}}\) and predicts multiple corrected anchor views.

\subsubsection{Block Context}

Besides the global raw-signal context, the module also computes a local block context from the anchor tensor itself. For each block, the raw anchor is averaged over channels or sensor groups and passed through a small MLP:
\[
h^{(m)}_{\mathrm{blk}} = f_{\mathrm{blk}}\!\left(Z^{(m)}_{\mathrm{raw}}\right)
\in \mathbb{R}^{B \times N_m \times d_b}.
\]
The correction network is then conditioned on the concatenated context
\[
h^{(m)}_{\mathrm{ctx}} = [h_{\mathrm{global}}; h^{(m)}_{\mathrm{blk}}].
\]

\subsubsection{Per-Family Affine Correction}

Let \(K\) denote the total number of anchor views. View \(0\) is always the identity view, which keeps the raw anchor unchanged. The remaining \(K-1\) views are produced by correction heads. Each correction head predicts, for every channel and TSF family, a scale parameter, a bias parameter, and a gating parameter. These family-level parameters are then broadcast to all feature dimensions within the corresponding family.

For correction head \(k\), the bounded affine transform is
\[
\widetilde{Z}^{(m)}_k
=
Z^{(m)}_{\mathrm{raw}}
\odot
\left(1 + s_{\max}\tanh(\hat{s}^{(m)}_k)\right)
+
b_{\max}\tanh(\hat{b}^{(m)}_k),
\]
where \(\hat{s}^{(m)}_k\) and \(\hat{b}^{(m)}_k\) are unconstrained outputs of the correction MLP, and \(s_{\max}\), \(b_{\max}\) bound the correction magnitude.

Instead of fully replacing the raw anchor, TCNet interpolates between the identity anchor and the corrected anchor:
\[
\lambda^{(m)}_k
=
\sigma(\alpha_k)\,\sigma(\hat{\lambda}^{(m)}_k),
\]
\[
Z^{(m)}_k
=
(1-\lambda^{(m)}_k)\odot Z^{(m)}_{\mathrm{raw}}
+
\lambda^{(m)}_k \odot \widetilde{Z}^{(m)}_k.
\]
This residual interpolation is central to the anchor formulation: the model can strengthen, weaken, or slightly shift a handcrafted feature family without discarding its original semantics.

The output of the correction module is therefore a multi-view anchor tensor
\[
Z^{(m)}_{\mathrm{multi}}
\in
\mathbb{R}^{B \times N_m \times (K C) \times D},
\]
formed by concatenating the identity view and all corrected views along the axis dimension.

\subsubsection{Regularization}

To prevent the correction module from overwriting the handcrafted anchors, TCNet regularizes the correction magnitude. The implementation penalizes the effective correction relative to the raw anchor, and optionally applies temporal smoothness regularization across adjacent blocks:
\[
\mathcal{L}_{\mathrm{delta}}
=
\sum_m \left\| Z^{(m)}_{\mathrm{corr}} - Z^{(m)}_{\mathrm{raw}} \right\|_1,
\]
\[
\mathcal{L}_{\mathrm{tv}}
=
\sum_m \sum_{n=2}^{N_m}
\left\|
\Delta^{(m)}_{n} - \Delta^{(m)}_{n-1}
\right\|_1,
\]
where \(\Delta^{(m)}\) denotes the blockwise correction. The total training loss is
\[
\mathcal{L}
=
\mathcal{L}_{\mathrm{cls}}
+
\alpha \mathcal{L}_{\mathrm{delta}}
+
\beta \mathcal{L}_{\mathrm{tv}}.
\]

\subsection{Hierarchical Fusion of Views, Sensors, Blocks, and Scales}

After correction, TCNet converts the multi-view anchor tensor into a fixed-size representation through a lightweight hierarchical fusion head.

\subsubsection{TSF-Group Projection}

The corrected anchor dimensions are first partitioned according to TSF family boundaries (e.g., filterbank, spectral, statistics, temporal crossing, quantiles, autocorrelation). Each family is projected independently into a shared latent space and concatenated:
\[
H^{(m)}_{\mathrm{proj}} \in \mathbb{R}^{B \times N_m \times (KC) \times d}.
\]
This preserves the structure of the handcrafted anchor families while allowing a compact learned re-embedding.

\subsubsection{View and Sensor-Group Fusion}

The projected tensor is reshaped into view and channel dimensions:
\[
H^{(m)}_{\mathrm{proj}} \rightarrow \mathbb{R}^{B \times N_m \times K \times C \times d}.
\]
For each predefined sensor group, TCNet first averages channels within the group and applies attention over the \(K\) anchor views. This yields one fused representation per sensor group. It then applies a second attention layer over sensor groups to obtain one block representation:
\[
H^{(m)}_{\mathrm{blk}} \in \mathbb{R}^{B \times N_m \times d}.
\]
This design explicitly separates three sources of variation: which corrected view is most useful, which sensor group is most informative, and which temporal blocks matter most.

\subsubsection{Block Pooling and Multi-Scale Fusion}

For each scale \(m\), TCNet computes a learned scalar score for each temporal block and performs attention pooling:
\[
h^{(m)}
=
\sum_{n=1}^{N_m}
a^{(m)}_n H^{(m)}_{\mathrm{blk},n},
\qquad
a^{(m)}_n = \mathrm{softmax}(g(H^{(m)}_{\mathrm{blk},n})).
\]
The resulting per-scale vectors are concatenated and linearly projected:
\[
h = \phi\left(\mathrm{Concat}\left(h^{(1)}, \dots, h^{(M)}\right)\right).
\]
Finally, a lightweight residual MLP classifier maps \(h\) to the class logits.

\subsection{Interpretation of the Architecture}

The final model can be viewed as a structured correction network over handcrafted motion statistics. The differentiable TSF extractor provides an explicit and semantically meaningful statistical scaffold. The temporal and frequency branches do not attempt to replace this scaffold; instead, they provide the context needed to determine when a feature family should be preserved, amplified, attenuated, or slightly shifted. The classifier then operates on these corrected anchors rather than on opaque latent activations.

This structure is what distinguishes TCNet from both classical handcrafted pipelines and standard deep HAR models. Compared with fixed TSF pipelines, TCNet can adapt anchor reliability to the current input. Compared with end-to-end latent encoders, it preserves explicit statistical structure throughout the forward pass.

\section{Experiments}
\label{sec:experiments}

This section addresses three questions. First, does TCNet outperform strong HAR baselines on standard benchmarks? Second, do the gains come from anchor adaptation rather than from additional branches or capacity alone? Third, what evidence shows that explicit anchors preserve useful statistical structure during learning? We answer these questions in the main experimental subsections and then report an extended transfer analysis on anchor-aligned pretraining.

\subsection{Datasets}
We evaluate TCNet on five widely used HAR benchmarks: USC-HAD~\cite{zhang2012usc}, UCI-HAR~\cite{anguita2013public}, Daphnet~\cite{bachlin2009wearable}, MHealth~\cite{banos2014mhealthdroid}, and PAMAP2~\cite{reiss2012introducing}. Together, these datasets cover a broad range of activity-recognition settings, including daily activity recognition, multi-location wearable sensing, and clinical motion monitoring. They vary substantially in activity granularity, sensor placement, number of channels, and class balance, providing a diverse test bed for evaluating whether feature anchors remain effective across different sensing conditions and task difficulties. For all datasets, we follow subject-independent train/test protocols, so that test subjects are never seen during training.

\textbf{USC-HAD~\cite{zhang2012usc}.}
USC-HAD is a benchmark for daily activity recognition collected from 14 subjects performing 12 activities, including locomotion and posture-related actions. The sensing unit is worn on the body and provides 3-axis accelerometer and 3-axis gyroscope measurements, yielding 6 IMU channels in total. The signals are sampled at 100\,Hz. Following the standard subject-independent protocol used in our experiments, subjects 1--12 are used for training and subjects 13--14 for testing. We segment the data into 200-sample windows, corresponding to 2 seconds of motion.

\textbf{UCI-HAR~\cite{anguita2013public}.}
UCI-HAR contains recordings from 30 volunteers performing 6 daily activities while carrying a smartphone on the waist. The activities include walking, walking upstairs, walking downstairs, sitting, standing, and laying. The smartphone provides accelerometer and gyroscope measurements sampled at 50\,Hz. In our experiments, we use the standard processed 9-channel version, which includes body acceleration, gyroscope, and total acceleration signals. We follow the standard subject-independent split with 23 subjects for training and 7 subjects for testing, and we use 128-sample windows, corresponding to 2.56 seconds.

\textbf{Daphnet~\cite{bachlin2009wearable}.}
The Daphnet Freezing of Gait dataset targets a clinically important binary recognition problem: detecting freezing episodes in Parkinson's disease patients. It contains recordings from 10 patients wearing accelerometers on the shank, thigh, and lower back. These three sensors provide 9 acceleration channels in total and are sampled at 64\,Hz. We use the standard binary labels (\textit{freezing} versus \textit{not freezing}). This dataset is strongly class-imbalanced and substantially more challenging than standard daily activity benchmarks because freezing events are sparse and patient-specific. Following the subject-independent protocol adopted in prior work, we train on subjects S01, S03, S06--S10, S02R02, S03R03, and S05R01, and test on S02R01, S04, and S05R02. We use 192-sample windows, corresponding to 3 seconds.

\textbf{MHealth~\cite{banos2014mhealthdroid}.}
The MHealth dataset consists of body-motion recordings from 10 subjects performing 12 physical activities. The sensing setup includes wearable devices placed on multiple body locations, which makes the dataset useful for evaluating cross-location motion patterns. In our configuration, we use acceleration signals from the chest, left ankle, and right lower arm together with gyroscope signals from the left ankle and right lower arm, resulting in 15 channels sampled at 50\,Hz. We follow a subject-independent split with subjects 2, 4, 5, 7, 8, 9, and 10 for training and subjects 1, 3, and 6 for testing. Input windows are 100 samples long, corresponding to 2 seconds.

\textbf{PAMAP2~\cite{reiss2012introducing}.}
PAMAP2 is the most complex benchmark in our evaluation. It records physical activities from 9 subjects wearing three IMUs placed on the hand, chest, and ankle. The original dataset includes accelerometer, gyroscope, magnetometer, temperature, and heart-rate measurements. Following common HAR preprocessing practice, we retain the IMU channels and use a 36-channel input formed from the three body locations. The signals are sampled at 100\,Hz. We follow the standard subject-independent split used in our experiments, training on subjects 101--105 and 107--109 and testing on subject 106. We use 256-sample windows, corresponding to 2.56 seconds.

Table~\ref{tab:protocols} provides a compact summary of the corresponding experimental configurations.

\begin{table*}[t]
\caption{Summary of the five HAR benchmarks used in our evaluation. The datasets span daily activity recognition, clinical freezing-of-gait detection, and multi-IMU physical activity monitoring, with 2 to 12 activity classes and 6 to 36 sensor channels. All experiments follow subject-independent train/test splits.}
  \label{tab:protocols}
  \small
  \setlength{\tabcolsep}{3pt}
  \begin{adjustbox}{max width=\linewidth}
  \begin{tabular}{lcccc}
    \toprule
    \textbf{Dataset} & \textbf{Window Size} & \textbf{Sampling Rate} & \textbf{Sensors (Channels)} & \textbf{Train / Test Subjects} \\
    \midrule
    USC-HAD~\cite{zhang2012usc} & 200 (2s) & 100Hz & Acc, Gyro (6) & 1-12 / 13,14 \\
    UCI-HAR~\cite{anguita2013public} & 128 (2.56s) & 50Hz & Acc, Gyro, TotalAcc (9) & 23 subjects / 7 subjects \\
    Daphnet~\cite{bachlin2009wearable} & 192 (3s) & 64Hz & 3x Acc (9) & S01,03,06-10, S02R02,03R03,05R01 / S02R01,04,05R02 \\
    MHealth~\cite{banos2014mhealthdroid} & 100 (2s) & 50Hz & Acc, Gyro, Mag (15) & 2, 4,5,7-10 / 1,3,6 \\
    PAMAP2~\cite{reiss2012introducing} & 256 (2.56s) & 100Hz & 3x IMU (36) & 101-105,107-109 / 106 \\
    \bottomrule
  \end{tabular}
  \end{adjustbox}
\end{table*}

\subsection{Implementation Details}
\label{sec:implementation}
For all datasets, we adopt a two-scale window strategy, where small blocks capture short-term activity transitions and large blocks provide longer activity-level context. The specific block sizes are dataset-dependent, as different benchmarks exhibit varying window sizes and sampling rates (see Table~\ref{tab:protocols}). In particular, the window size plays a primary role in determining our configuration. Accordingly, for USC-HAD we use block sizes of 50 (small) and 100 (large); for UCI-HAR, 32 and 128; for Daphnet, 32 and 192; for MHealth, 32 and 64; and for PAMAP2, 32 and 128. Both scales are processed independently and fused before final classification. We use a 4-head adaptation module ($K{=}4$) for all datasets except USC-HAD ($K{=}2$); the per-group projection dimension is $D_{proj} = 128$ and the context dimension is $D_{ctx} = 256$.
\textbf{Context branches}: The MainTime branch uses parallel 1D convolutions with kernel sizes [5, 11, 21]; the MainFreq branch applies STFT with FFT sizes [32, 64, 128]. Both branches output $D_{ctx}$-dimensional vectors after MLP aggregation.
\textbf{Training}: We optimize with Adam (weight decay $10^{-4}$) and a cosine annealing scheduler with early stopping (patience 20 epochs). Learning rates are tuned per dataset: $10^{-3}$ for Daphnet, MHealth, and PAMAP2; $5{\times}10^{-4}$ for UCI-HAR; and $3{\times}10^{-4}$ for USC-HAD. Maximum training budgets are 300 epochs for Daphnet, UCI-HAR, and USC-HAD; 120 for PAMAP2; and 50 for MHealth. The correction regularization weights are fixed uniformly at $\alpha{=}\beta{=}10^{-4}$ across all five datasets.

\noindent\textbf{Extended-analysis models for transfer experiments.}
Compact TCNet uses a single time-domain branch (1D convolution, kernel size 7, 2 layers, $d_\mathrm{time}{=}64$), a single frequency branch (STFT with $n_\mathrm{fft}{=}64$, $d_\mathrm{freq}{=}64$), and one correction expert ($K_\mathrm{corr}{=}1$) with block context dimension 16. A 128-dimensional content head fuses the branch outputs and the corrected anchor feature; the final frozen representation is the 256-dimensional concatenation of the content embedding and the pooled corrected TSF. The SSL block size is 32 samples for UCI-HAR, Daphnet, MHealth, and PAMAP2 (50 samples for USC-HAD). Pretraining uses Adam ($lr{=}10^{-3}$, weight decay $10^{-4}$) for 20 epochs with batch size 256. Compact TCNet has approximately \textbf{0.14\,M} trainable parameters (identical across datasets).

The UKB-SSL encoder (ssl-wearables ResNet~\cite{yuan2022selfsupervised}) comprises five progressive residual stages with channel widths [64, 128, 256, 512, 1024] and anti-aliased downsampling, yielding a 1024-dimensional representation, with approximately \textbf{10.46\,M} parameters. This encoder is kept fully frozen during all downstream evaluations; only the RF or MLP head is trained on target-domain labels.

\noindent To verify that TCNet's gains are not an artifact of model scale, Fig.~\ref{fig:param_efficiency} plots mean macro-F1 against trainable parameters (left) and per-cell margins over the strongest competitor (right). TCNet uses only $\sim$1/20 the parameters of rTsfNet (1.00\,M vs.\ 20.50\,M) yet achieves higher mean mF1 (88.28 vs.\ 83.01), and it leads on the majority of dataset--metric cells. For reference, Compact TCNet is a \textbf{0.14\,M} encoder used only for transfer analysis, while the UKB-SSL ResNet encoder has \textbf{10.46\,M} parameters but is fully frozen during downstream evaluation.

\subsection{Strong RF-TSF Baseline}
\label{sec:rf_baseline}

Most prior work compares deep HAR models against Random Forest baselines built on generic handcrafted features or library-default settings, which makes the comparison less stringent when the neural model uses a different feature space internally. To provide a fair handcrafted reference, we construct a strong RF-TSF baseline using the \emph{same multi-scale handcrafted feature families} that TCNet uses as feature anchors.

\noindent\textbf{Feature extraction.}
For each dataset, we apply the same two block sizes used by TCNet (Sec.~\ref{sec:implementation}) and partition each IMU window into non-overlapping temporal blocks at each scale. For every block and every sensor channel, we compute the same handcrafted TSF families used to form TCNet's anchors, including filterbank responses, spectral summaries, basic statistical descriptors, higher-order distributional shape descriptors, temporal crossing features, quantile features, and autocorrelation features. We then average these block-level features over time within each scale to obtain one fixed-length handcrafted representation per scale, and concatenate the two scale-level vectors into the final feature representation for the window.

This protocol ensures that the RF baseline receives essentially the same handcrafted statistical information as TCNet's anchor pathway, but without any learned context-conditioned correction, multi-view adaptation, or neural fusion. As a result, the RF-TSF baseline serves as a stringent reference for assessing whether TCNet's gains come from anchor adaptation rather than from the handcrafted feature set itself.

\noindent\textbf{Classifier and training protocol.}
We use a Random Forest classifier with 300 trees, maximum depth 20, balanced class weights, and a fixed random seed. No dataset-specific hyperparameter tuning is applied. Following the same evaluation protocol as TCNet, we train on the designated training subjects and report results on the held-out test subjects.
 This produces the \emph{RF-TSF} row in Table~\ref{tab:results}.

\subsection{Main Results}

\begin{table*}[t]
\caption{Main results across five HAR benchmarks. Macro-F1 and accuracy are reported for 17 methods spanning classical, deep, hybrid, and LLM-based approaches; per-column best in bold. For our method, we additionally report the relative improvement over the strongest competing method in each column (shown in {\color{red}red parentheses} ), to explicitly highlight the performance margin. TCNet leads on USC-HAD, Daphnet, MHealth, and PAMAP2 (mF1), with substantial margins over rTsfNet and deep baselines, and surpasses the RF-TSF ceiling that most deep methods fail to exceed.  See Fig.~\ref{fig:param_efficiency} (right) for per-cell margins.}
  \label{tab:results}
  \footnotesize
  \setlength{\tabcolsep}{3.5pt}
  \begin{adjustbox}{max width=\linewidth}
  \begin{tabular}{l|cc|cc|cc|cc|cc}
    \toprule
    \textbf{Method} & \multicolumn{2}{c|}{\textbf{USC-HAD}} & \multicolumn{2}{c|}{\textbf{UCI-HAR}} & \multicolumn{2}{c|}{\textbf{Daphnet}} & \multicolumn{2}{c|}{\textbf{MHealth}} & \multicolumn{2}{c}{\textbf{PAMAP2}} \\
     & mF1 & Acc & mF1 & Acc & mF1 & Acc & mF1 & Acc & mF1 & Acc \\
    \midrule
    RandomForest & 57.61 & 67.39 & 92.57 & 92.67 & 76.90 & 96.10 & 89.56 & 88.92 & 89.39 & 89.92 \\
    \midrule
    DeepConvLSTM~\cite{deepconvlstm2016} (2016) & 48.80 & 50.60 & 89.20 & 89.20 & 48.55 & 94.35 & 75.00 & 76.00 & 78.40 & 78.20 \\
    Transformer~\cite{vaswani2017attention} (2017) & 49.21 & 54.42 & 90.82 & 90.84 & 76.67 & 96.10 & 93.35 & 93.01 & 85.29 & 86.96 \\
    Reformer~\cite{kitaev2020reformer} (2020) & 40.33 & 45.69 & 92.55 & 92.50 & 71.81 & 95.60 & 92.19 & 91.71 & 85.22 & 86.81 \\
    Informer~\cite{zhou2021informer} (2021) & 46.29 & 53.67 & 92.35 & 92.40 & 73.90 & 95.78 & 90.06 & 89.47 & 86.25 & 87.32 \\
    FEDformer~\cite{zhou2022fedformer} (2022) & 51.37 & 62.91 & 90.83 & 90.94 & 73.51 & 95.74 & 91.85 & 91.51 & 81.92 & 83.48 \\
    FiLM~\cite{zhou2022film} (2022) & 55.66 & 64.23 & 87.71 & 88.06 & 72.64 & 95.02 & 88.22 & 88.27 & 86.05 & 86.81 \\
    LightTS~\cite{zhang2022lightts} (2022) & 59.01 & 67.78 & 90.89 & 91.18 & 61.05 & 94.71 & 90.85 & 90.91 & 89.05 & 89.14 \\
    Pyraformer~\cite{liu2022pyraformer} (2022) & 46.26 & 51.20 & 91.09 & 91.11 & 80.34 & 96.23 & 90.75 & 90.51 & 86.05 & 87.01 \\
    Crossformer~\cite{zhang2023crossformer} (2023) & 57.23 & 67.21 & 90.83 & 90.87 & 71.10 & 95.33 & 81.39 & 81.38 & 91.29 & 91.84 \\
    TimesNet~\cite{wu2023timesnet} (2023) & 54.19 & 60.79 & 92.67 & 92.67 & 70.58 & 94.71 & 90.89 & 91.21 & 85.97 & 86.23 \\
    iTransformer~\cite{liu2024itransformer} (2024) & 52.63 & 58.90 & 93.37 & 93.35 & 68.09 & 95.15 & 86.01 & 85.32 & 89.46 & 90.18 \\
    ICGNet~\cite{dua2023inception} (2023) &46.93 &52.00 & 92.62 & 92.47 & 56.16 & 91.88 &89.73 &90.11 &85.36 & 89.39 \\
    mobileHART~\cite{ek2023transformer} (2023) &54.54 &59.43 & 94.69 & 94.71 & 48.55 & 94.57 &64.93 &63.31 &90.67 & 93.47 \\
    MchCnnGRU~\cite{lu2022multichannel} (2022) &43.07 &52.60 & 95.49 & 95.55 & 48.55 & 94.35 &89.38 &89.72 &91.79 & 94.38 \\
    SensorLLM~\cite{sensorllm2025} (2025) & 61.20 & 62.60 & 91.20 & 90.80 & - & - & 89.40 & 89.00 & 86.20 & 87.20 \\
    rTsfNet~\cite{rTsfNet2024} (2024) & 65.76 & 64.85 & \textbf{97.79} & \textbf{97.76} & 70.51 & 95.65 & 87.46 & 89.75 & 93.53 & \textbf{95.35} \\
    \midrule
\shortstack{\textbf{TCNet} (Ours)}
& \shortstack{\textbf{70.21}\\ {\color{red}(+4.45)}}
& \shortstack{\textbf{76.21}\\ {\color{red}(+8.43)}}
& \shortstack{97.65\\ {(-0.14)}}
& \shortstack{97.59\\ {(-0.17)}}
& \shortstack{\textbf{85.12}\\ {\color{red}(+4.78)}}
& \shortstack{\textbf{97.04}\\ {\color{red}(+0.81)}}
& \shortstack{\textbf{93.91}\\ {\color{red}(+0.56)}}
& \shortstack{\textbf{93.71}\\ {\color{red}(+0.70)}}
& \shortstack{\textbf{94.52}\\ {\color{red}(+0.99)}}
& \shortstack{95.22\\ {(-0.13)}} \\
    \bottomrule
  \end{tabular}
  \end{adjustbox}
\end{table*}

\begin{figure}[htb]
  \centering
  \includegraphics[width=\linewidth]{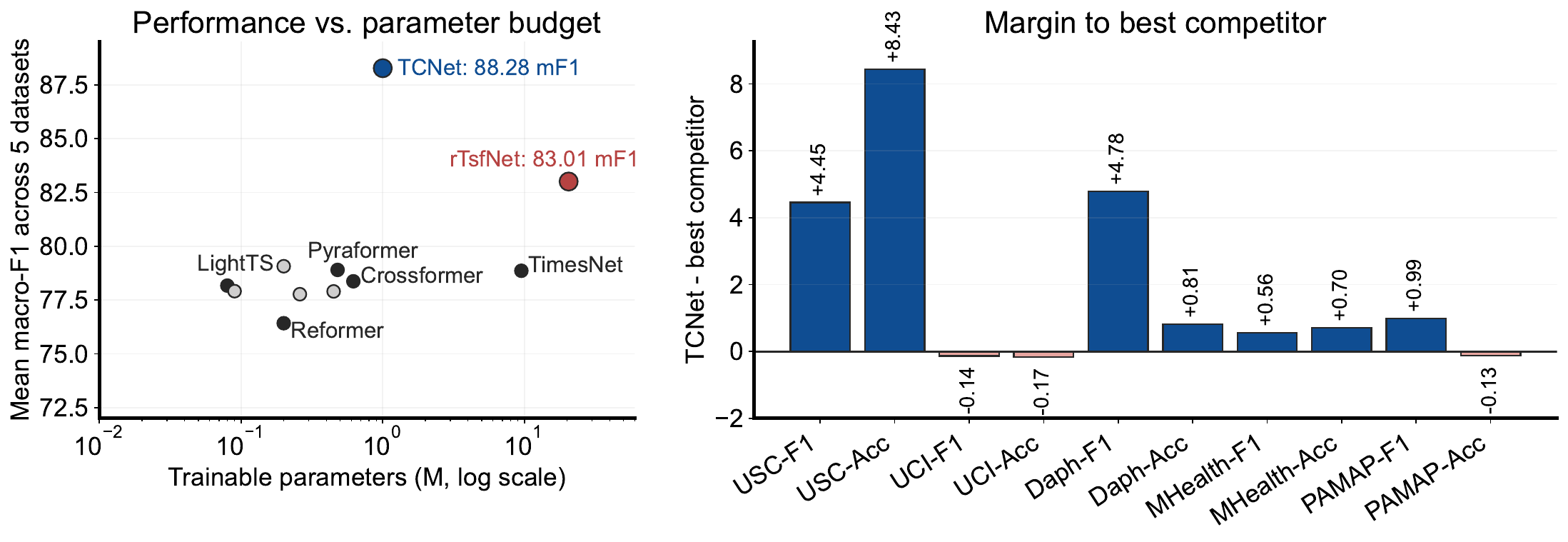}
  \caption{Parameter efficiency and per-cell performance margins. \textbf{Left:} Mean macro-F1 versus trainable parameters (log scale). TCNet uses approximately 1/20 the parameters of rTsfNet (1.00\,M vs.\ 20.50\,M) while achieving higher mean mF1 (88.28 vs.\ 83.01). \textbf{Right:} Per-cell margin relative to the strongest competing method in Table~\ref{tab:results}; positive values (blue) indicate cells where TCNet leads. TCNet wins on most dataset--metric combinations, and the few negative cells have margins under 1 point.}
  \label{fig:param_efficiency}
\end{figure}

Table~\ref{tab:results} compares TCNet against a comprehensive set of baselines spanning classical methods (Random Forest on TSFs), deep time-series architectures (generic sequence models such as Transformer~\cite{vaswani2017attention}, Reformer~\cite{kitaev2020reformer}, Informer~\cite{zhou2021informer}, FEDformer~\cite{zhou2022fedformer}, FiLM~\cite{zhou2022film}, LightTS~\cite{zhang2022lightts}, Pyraformer~\cite{liu2022pyraformer}, Crossformer~\cite{zhang2023crossformer}, DLinear~\cite{zeng2023dlinear}, MICN~\cite{wang2023micn}, PatchTST~\cite{nie2023patchtst}, TimesNet~\cite{wu2023timesnet}, and iTransformer~\cite{liu2024itransformer}; recurrent and CNN-based HAR models such as DeepConvLSTM~\cite{deepconvlstm2016}; HAR-specific CNN--GRU hybrids such as ICGNet~\cite{dua2023inception} and MchCnnGRU~\cite{lu2022multichannel}; and the Transformer-based HAR model mobileHART~\cite{ek2023transformer}), hybrid methods (rTsfNet~\cite{rTsfNet2024}), and LLM-based approaches (SensorLLM~\cite{sensorllm2025}).

Table~\ref{tab:results} shows that TCNet achieves the highest mF1 on four of the five benchmarks. The largest gains appear on USC-HAD (+4.5 over rTsfNet), Daphnet (+14.6), and MHealth (+6.5), where activity statistics vary substantially across subjects. On UCI-HAR, rTsfNet retains a slight advantage in both metrics. On PAMAP2, TCNet leads in mF1 but trails rTsfNet by 0.13 points in accuracy. The per-cell summary in Fig.~\ref{fig:param_efficiency} (right) shows that TCNet leads on most dataset--metric cells, and the few losses are all within one point. This pattern suggests that context-conditioned anchor adaptation is especially helpful when inter-subject variability is high.

\subsection{Which Components Drive the Gain?}
\label{sec:ablation}

\begin{figure*}[t]
  \centering
  \includegraphics[width=\linewidth]{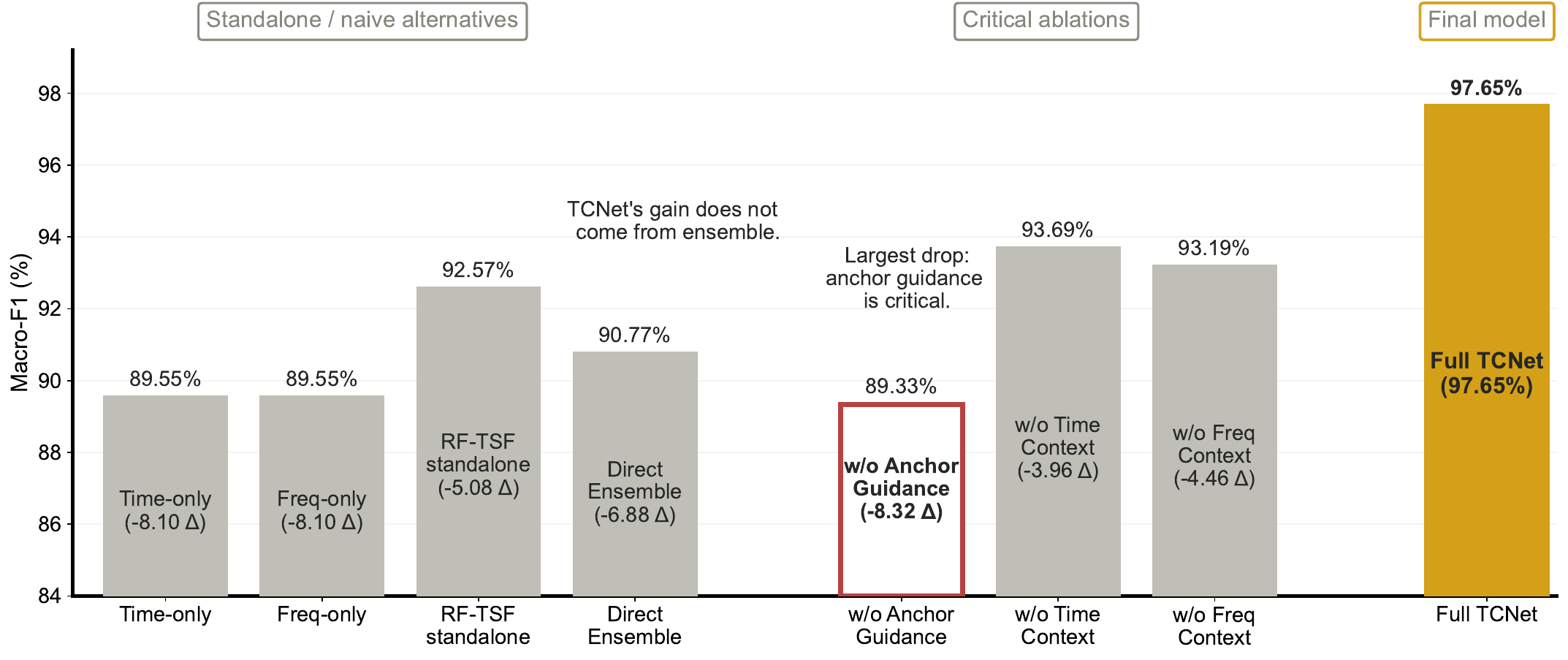}
  \caption{Ablation evidence that TCNet's gain comes primarily from anchor adaptation rather than branch fusion alone. \textbf{Left:} The standalone time branch, standalone frequency branch, RF-TSF baseline, and their direct ensemble all remain substantially below Full TCNet on UCI-HAR. \textbf{Right:} Removing anchor correction causes the largest single performance drop ($-8.32$ mF1), larger than removing either the time-context or frequency-context branch. Moreover, the no-anchor-correction variant falls below the standalone RF-TSF baseline, indicating that TCNet's advantage comes from the interaction between raw-signal context and explicit handcrafted anchors, not from added capacity alone.}
  \label{fig:component_gap_summary}
\end{figure*}

The main results establish that TCNet is competitive, but our central claim is more specific: its improvement comes from \emph{adapting explicit feature anchors}, rather than from simply combining multiple branches or increasing model capacity. Figure~\ref{fig:component_gap_summary} tests this claim directly on UCI-HAR.

Figure~\ref{fig:component_gap_summary} (left) first shows that the gain cannot be explained by naive combination of heterogeneous predictors. The standalone time branch and standalone frequency branch each perform far below Full TCNet, and even a direct ensemble of the time branch, frequency branch, and RF-TSF predictor remains 6.88 mF1 points below the full model. Notably, this direct ensemble also underperforms the standalone RF-TSF baseline (92.57\% mF1). This rules out a simple ``more branches is better'' explanation.

Figure~\ref{fig:component_gap_summary} (right) then isolates the contribution of each component within TCNet. Removing the time-domain context branch reduces mF1 by 3.96 points, while removing the frequency-domain context branch reduces it by 4.46 points, showing that both forms of raw-signal context are useful and complementary. 

The largest single drop, however, comes from removing anchor correction, which lowers mF1 by 8.32 points. In this setting, performance falls to 89.33\%, below the standalone RF-TSF baseline. This result is particularly important: once context-conditioned correction is removed, the model no longer improves over a strong handcrafted pipeline built on the same underlying TSF space. The gain of TCNet therefore does not come from extra branches or additional parameters by themselves; it comes from using raw-signal context to adapt explicit handcrafted anchors in feature space.

\subsection{Raw versus Corrected Feature Distributions}
\label{sec:analysis}

\begin{figure}[t]
  \centering
\includegraphics[width=\linewidth]{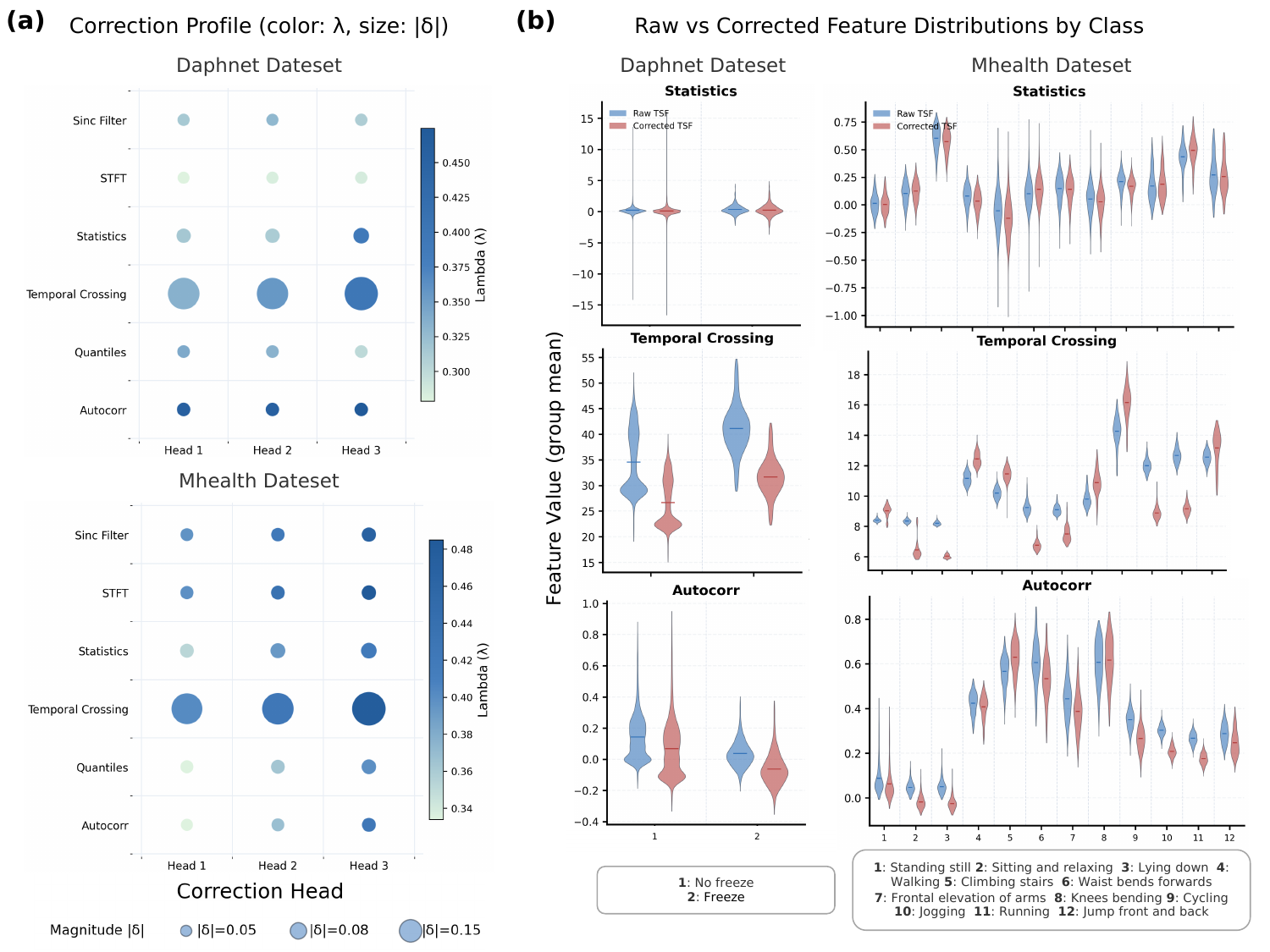}
  \caption{Raw versus anchor-guided feature distributions for three TSF families on Daphnet (top) and MHealth (bottom). Violin plots compare raw (blue) and adapted (red) distributions. \textbf{Statistics} changes minimally in both datasets. \textbf{Temporal Crossing} undergoes the strongest modulation: in MHealth, dynamic activities (Jogging, Running) are downscaled more than static ones, sharpening class separation while preserving the relative ordering. \textbf{Autocorr} distributions narrow slightly while maintaining the semantic class structure. The consistent pattern across families is selective tightening or rescaling rather than erasure, indicating that adapted anchors remain interpretable.}
  \label{fig:raw_adapted}
\end{figure}

To understand what the correction module actually learns, we compare selected TSF families before and after context-conditioned correction. Figure~\ref{fig:raw_adapted} examines  anchor families on two datasets with very different label structure: Daphnet (binary freezing-of-gait detection) and MHealth (12-class activity recognition). The main pattern is consistent across both settings: correction is \emph{selective rather than uniform}. Some anchor families are left nearly unchanged, whereas others are systematically rescaled or tightened in ways that improve class separation while preserving their semantic interpretation.

Panel (a) summarizes this pattern at the family level. Across both datasets, \textbf{Temporal Crossing} receives the largest correction magnitude, indicating that the model relies most heavily on adjusting crossing-based dynamics. By contrast, \textbf{Statistics} and several other families receive noticeably smaller corrections, suggesting that these anchors are already reasonably aligned with the downstream task and therefore need less modification. This family-dependent correction profile is consistent with the role of TCNet: it does not rewrite all handcrafted features indiscriminately, but instead applies stronger updates where the raw anchors are less discriminative.

Panel (b) shows the corresponding raw-versus-corrected feature distributions by class. On \textbf{Daphnet}, the correction primarily sharpens the binary separation between \textit{No freeze} and \textit{Freeze}. \textbf{Statistics} remains nearly unchanged for both classes, indicating that this family is largely preserved. \textbf{Temporal Crossing} shows the clearest correction effect: the raw distributions are broad, whereas the corrected distributions become tighter and more separated, especially along the class-relevant direction. \textbf{Autocorr} is corrected more mildly; its tails are compressed and its spread is reduced without destroying the overall class structure.

The same pattern appears on \textbf{MHealth}, but in a more fine-grained multi-class setting. \textbf{Statistics} again changes little, with raw and corrected distributions largely overlapping across activities. In contrast, \textbf{Temporal Crossing} is corrected substantially: low-dynamic activities such as standing, sitting, and lying remain in the lower range, whereas more dynamic activities such as walking, climbing stairs, jogging, and running remain in the higher range, but with improved separation after correction. Importantly, the correction sharpens class structure without scrambling the original ordering by activity intensity. \textbf{Autocorr} shows a smaller but still meaningful adjustment: periodic activities retain relatively higher values, while the corrected distributions become narrower and more class-consistent.

Taken together, Fig.~\ref{fig:raw_adapted} supports a specific interpretation of TCNet's behavior. The model does not improve performance by erasing handcrafted anchors and replacing them with opaque latent features. Instead, it performs \emph{structure-preserving correction}: weak or overly broad anchor families are selectively tightened, shifted, or rescaled, whereas already useful anchors are left largely intact. This is precisely the behavior expected from feature anchors---explicit statistical representations that remain interpretable while being refined by neural context.

\subsection{Do Deep Networks Implicitly Learn Handcrafted Feature Structure?}
\label{sec:linear_probe}

The gap between RF-TSF and end-to-end deep models in Table~\ref{tab:results} does not, by itself, reveal what a neural encoder represents internally. We therefore test a more direct representation question: given a frozen encoder, how well can a linear model predict each TSF family from the encoder embedding?

\textbf{Protocol.}
We use TimesNet~\cite{wu2023timesnet} trained to convergence on each HAR dataset as the encoder ($92.7\%$ mF1 on UCI-HAR, $70.6\%$ on Daphnet, and $86.0\%$ on PAMAP2). We extract the pre-projection representation, compute ground-truth TSF values for the same windows with our differentiable extractor, and group them into six families: Filterbank (Sinc), Spectral (STFT), Statistics, Temporal Crossing, Quantiles, and Autocorrelation. For each family, we train a Ridge regression from the frozen embeddings to the TSF values on the training split and report $R^2$ on the held-out test split across all five benchmark datasets. A TimesNet with the same architecture but random initialization serves as a control.

\begin{figure}[t]
  \centering
  \includegraphics[width=\linewidth]{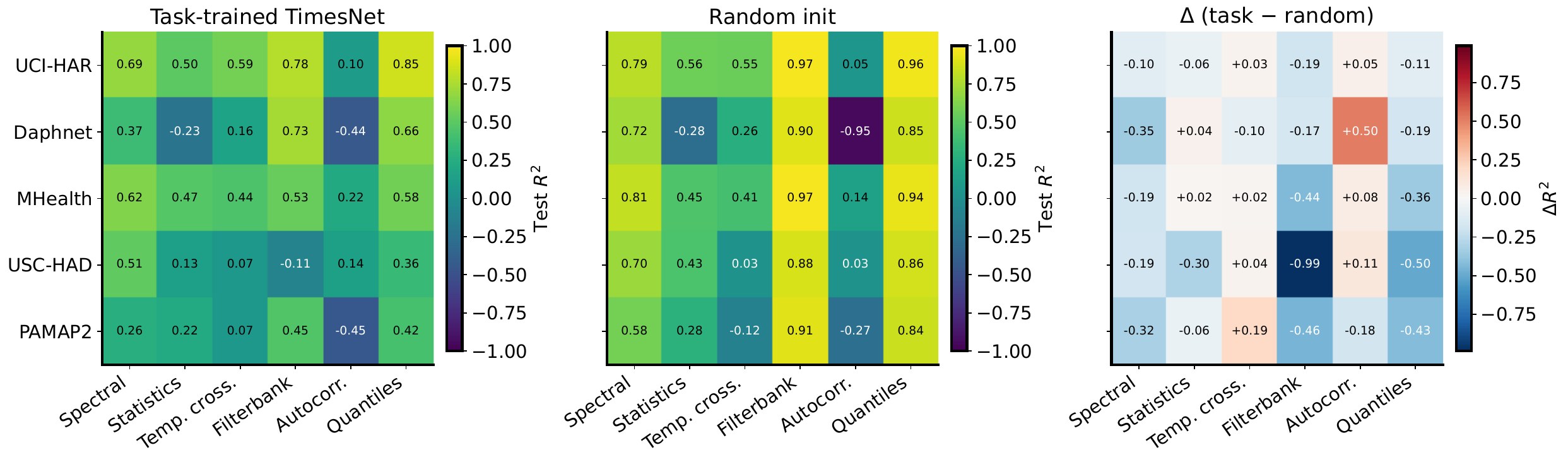}
  \caption{Linear probe $R^2$ from frozen TimesNet embeddings to individual TSF families across five HAR datasets. \textbf{Left:} task-trained encoder. \textbf{Center:} random-initialization control with the same architecture and no training. \textbf{Right:} $\Delta R^2 = R^2_{\mathrm{task}} - R^2_{\mathrm{random}}$; red cells indicate families whose linear accessibility \emph{decreases} after supervised training. Three families, Autocorr ($\overline{R^2} = -0.09$), Statistics ($\overline{R^2} = 0.22$), and Temporal Crossing ($\overline{R^2} = 0.27$), remain weakly encoded despite high classifier accuracy, while amplitude-correlated families (Filterbank and Quantiles) are suppressed by training. This pattern motivates keeping discriminative TSFs explicit as anchors.}
  \label{fig:linear_probe_r2}
\end{figure}

\begin{figure}[t]
  \centering
  \includegraphics[width=\linewidth]{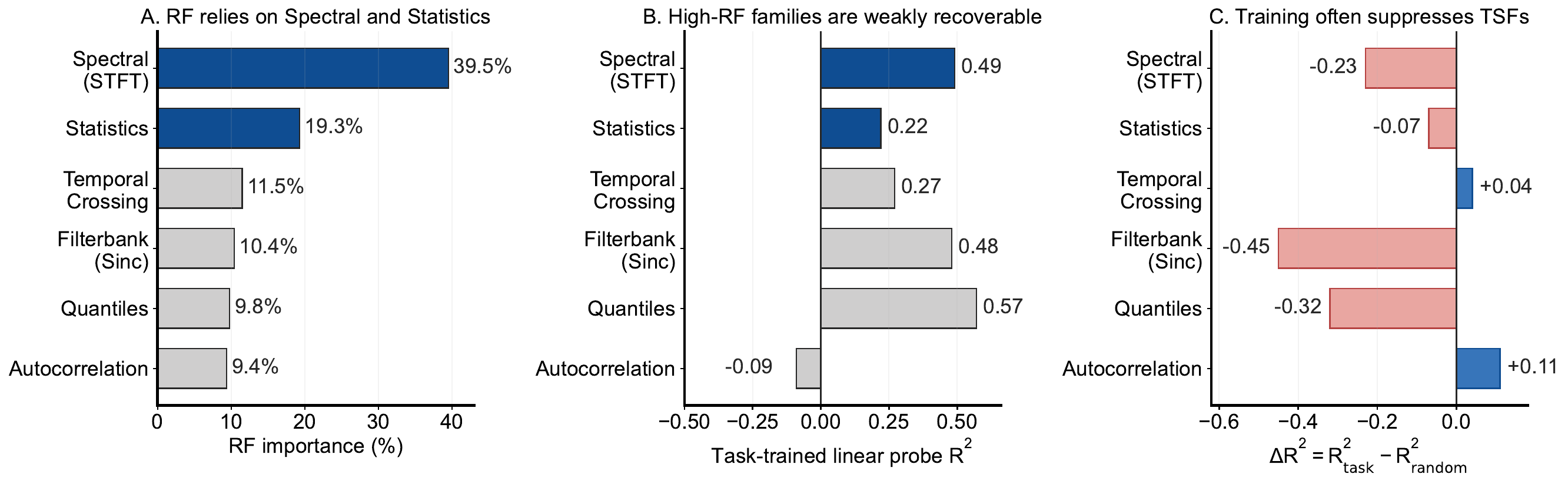}
  \caption{Structural mismatch between TSF importance and latent encodability. \textbf{Panel A:} RF feature importance identifies Spectral (39.5\%) and Statistics (19.3\%) as the dominant discriminative families. \textbf{Panel B:} Linear probe $R^2$ on the task-trained TimesNet encoder shows that these same families are only weakly recoverable (Spectral: 0.49; Statistics: 0.22; Autocorr: $<$0). \textbf{Panel C:} $\Delta R^2$ confirms that supervised training often suppresses rather than restores TSF accessibility, particularly for Filterbank and Quantiles. This mismatch between high RF importance and low probe $R^2$ motivates keeping discriminative TSF families explicit as anchors.}
  \label{fig:linear_probe_anchor_summary}
\end{figure}

\textbf{Results.}
Figure~\ref{fig:linear_probe_r2} presents the full $R^2$ matrix across TSF families and datasets. Figure~\ref{fig:linear_probe_anchor_summary} summarizes the main argument in three panels: RF feature importance (Panel A), probe $R^2$ on the task-trained encoder (Panel B), and the change relative to the random-initialization control (Panel C). Two patterns appear consistently.

First, task training reduces the linear decodability of amplitude-correlated families. The randomly initialized TimesNet achieves $R^2 = 0.93$ for Filterbank and $0.89$ for Quantiles; training on HAR labels reduces these values to $0.48$ and $0.57$, respectively. This indicates that supervised optimization reorganizes activations toward task-separating directions at the expense of preserving these correlations.

Second, three families remain weakly encoded even after training: Autocorrelation ($R^2 = -0.09$), Statistics ($0.22$), and Temporal Crossing ($0.27$). These low values persist despite high classifier accuracy, which indicates that the encoder can be predictive without making these TSF families linearly accessible.

\textbf{Importance--encodability inversion.}
Panels A and B of Fig.~\ref{fig:linear_probe_anchor_summary} quantify the mismatch. Spectral, which accounts for 39.5\% of RF importance, achieves only $R^2 = 0.49$ in the trained encoder. Statistics, which accounts for 19.3\% of RF importance, reaches only $R^2 = 0.22$, and Autocorrelation has mean $R^2 < 0$. The TSF families that matter most to the strongest classical baseline are therefore not the ones that remain most accessible in latent space.

\textbf{Interpretation.}
Panel C helps separate architectural bias from task training. The random-initialization control achieves high $R^2$ for Filterbank (0.93) and Quantiles (0.89), which indicates that some amplitude-correlated structure is available from architecture alone. Supervised training shifts representations toward task-separating directions but does not recover Spectral or Statistics, and it suppresses several families relative to the control. Since Spectral and Statistics together account for nearly 60\% of RF importance yet have task-trained $R^2$ values of only 0.49 and 0.22, keeping these families explicit as anchors is a practical alternative to relying on implicit recovery in latent space.

\subsection{Extended Analysis: Anchor-Aligned Pretraining Under Transfer}
\label{sec:ssl_pretrain}

The results above support the main claim of the paper. We next ask a related transfer question: if explicit TSF structure matters, can anchor-aligned pretraining preserve that structure better than large-scale latent pretraining under domain shift? We design a controlled experiment to isolate these effects.

\textbf{Experimental design.}
We evaluate on UCI-HAR and PAMAP2 because they represent two complementary transfer settings. UCI-HAR is a single-device benchmark in which subjects wear one smartphone on the waist, which is closer to the \textit{ssl-wearables} source domain that uses a single wrist-worn Axivity AX3 accelerometer from the UK Biobank~\cite{yuan2022selfsupervised}. PAMAP2 is more complex: each sample comprises signals from three body-worn IMUs, yielding a 36-channel input after concatenation. The two datasets therefore test transfer in a relatively matched single-device regime and in a more demanding multi-device regime.

\textbf{Compact TCNet.}
To enable a fair self-supervised comparison, we introduce Compact TCNet, a simplified encoder derived from the full TCNet architecture. It retains the three core ingredients of TCNet: a lightweight time context branch ($h_{time}$), a lightweight frequency context branch ($h_{freq}$) based on a single-resolution STFT pipeline, and a differentiable TSF extractor ($TSF_{raw}$). We remove multi-scale blocks and the heavier downstream head so that frozen-feature evaluation is consistent across datasets. Each input is a single tri-axial sensor stream $x \in \mathbb{R}^{3 \times L}$. The anchor adaptation module produces a single adapted view, and the encoder outputs the pooled adapted anchor feature as the frozen representation:
\begin{equation}
TSF_{corr} = (1-\lambda) \odot TSF_{raw} + \lambda \odot \bigl(TSF_{raw} \odot (1+s) + b\bigr).
\end{equation}
This design preserves the anchor-guided inductive bias of TCNet while keeping the pretraining experiment lightweight and controlled.

For multi-sensor datasets, each HAR window is decomposed into independent SSL samples by treating every tri-axial sensor group as a separate pretraining instance; channels are partitioned into consecutive 3-axis groups $(0,1,2)$, $(3,4,5)$, and so on. During downstream evaluation, frozen features from all sensor streams are concatenated to form the final sample representation.

\textbf{Pretraining protocol.}
To avoid confounding from a different SSL objective, we adopt the same three pretext tasks used by Yuan et al.~\cite{yuan2022selfsupervised}: Arrow of Time (AoT), Permutation, and TimeWarp. AoT reverses the temporal order of the sequence; Permutation shuffles four equal-length chunks (minimum chunk length: 10 timestamps); and TimeWarp applies random local stretching and compression. Each task is cast as binary prediction (transformation applied vs.\ not), sampled independently with probability 0.5, and optimized with equal loss weights. The full pretraining objective is the mean of the three binary cross-entropy losses plus a lightweight regularizer that penalizes large deviations from the raw anchors. This alignment helps isolate the role of encoder architecture and anchor guidance rather than the SSL task family.

\begin{figure}[t]
  \centering
  \includegraphics[width=\linewidth]{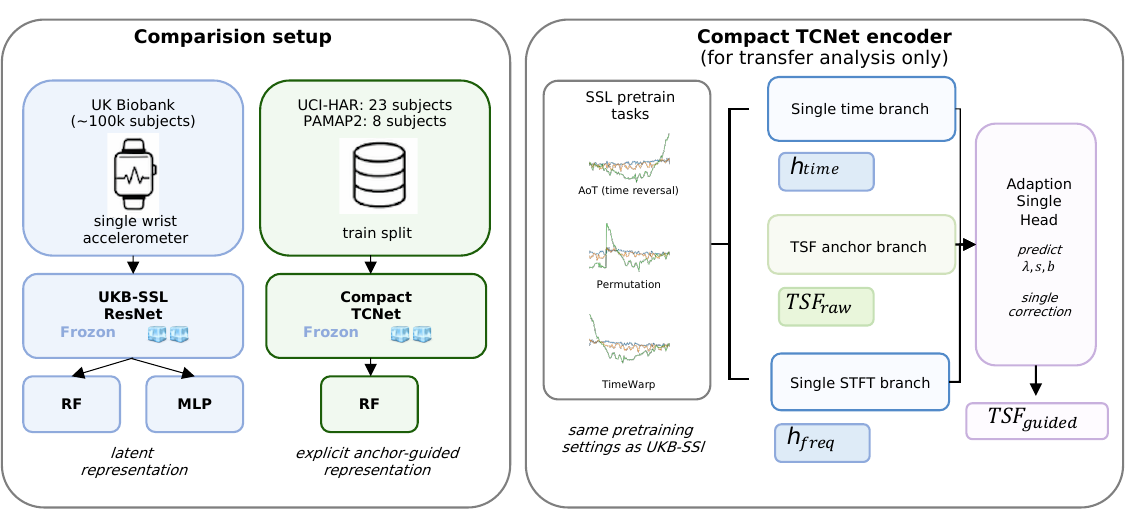}
  \caption{Compact TCNet pretraining pipeline. A lightweight TCNet encoder (0.14\,M parameters; single time branch, single frequency branch, and single-view anchor adaptation) is pretrained with three binary pretext tasks: Arrow of Time (temporal reversal detection), Permutation (chunk order detection), and TimeWarp (local stretch/compression detection), following Yuan et al.~\cite{yuan2022selfsupervised}. These tasks reward encoding statistical regularities such as energy profiles, crossing rates, and autocorrelation structure, which are also captured by handcrafted TSFs. After pretraining, the encoder is frozen and a Random Forest head is trained on target-domain labels.}
  \label{fig:pretrain_framework}
\end{figure}

\begin{figure*}[t]
  \centering
  \includegraphics[width=\linewidth]{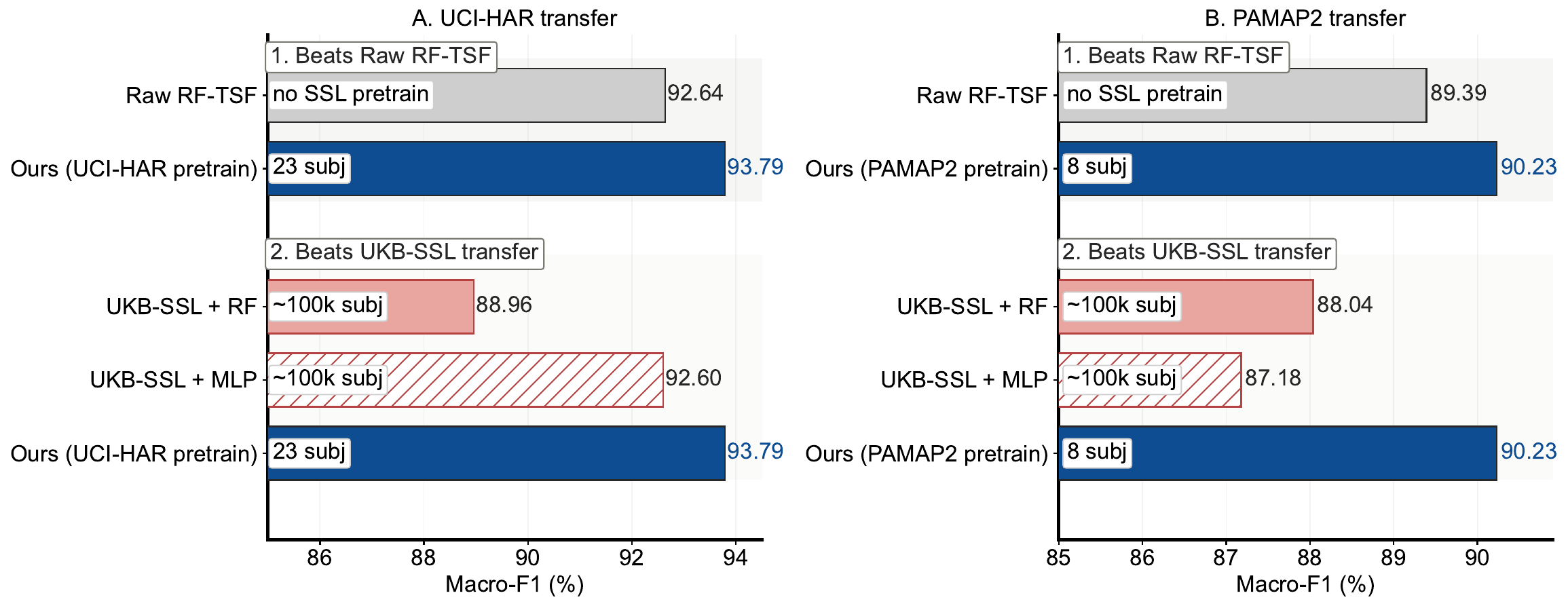}
  \caption{Compact TCNet pretrained on $\leq$23 subjects matches or surpasses UKB-SSL pretrained on ${\sim}$100k subjects. Panels A (UCI-HAR) and B (PAMAP2) each present two comparisons. \textbf{Left block:} Compact TCNet + RF improves over Raw RF-TSF (UCI-HAR: 93.79 vs.\ 92.64 mF1; PAMAP2: 90.23 vs.\ 89.39), confirming that anchor-guided pretraining adds value beyond static features. \textbf{Right block:} the same representation surpasses both UKB-SSL + RF and UKB-SSL + MLP despite using orders-of-magnitude less pretraining data, suggesting that in-domain pretext tasks aligned with TSF structure transfer more effectively than large-scale cross-domain pretraining under sensor mismatch.}
  \label{fig:ssl_pretrain_summary}
\end{figure*}

We compare five conditions on UCI-HAR and PAMAP2 (Table~\ref{tab:ssl_pretrain}), and summarize the main reviewer-facing takeaway visually in Fig.~\ref{fig:ssl_pretrain_summary}:

\begin{enumerate}
  \item \textbf{Raw RF-TSF}: a Random Forest trained on 70+ handcrafted features; the reference anchor-only baseline requiring no neural pretraining.
  \item \textbf{UKB-SSL + RF}: the \textit{ssl-wearables} ResNet encoder~\cite{yuan2022selfsupervised} pretrained on ${\sim}$100k UK Biobank subjects, evaluated with a frozen Random Forest.
  \item \textbf{UKB-SSL + MLP}: the same frozen encoder evaluated with a two-layer MLP head fine-tuned on the target dataset.
  \item \textbf{Random Init + MLP}: the same ResNet with \textit{random weights} (no pretraining), evaluated identically to UKB-SSL + MLP. This isolates architectural inductive bias from learned representations.
  \item \textbf{Compact TCNet + RF} (ours): Compact TCNet (Fig.~\ref{fig:pretrain_framework}) pretrained \textit{in-domain} on the target dataset's training split using the same three SSL tasks, then evaluated with a frozen Random Forest.
\end{enumerate}

\begin{table*}[t]
  \caption{Transfer comparison: anchor-guided in-domain pretraining versus large-scale cross-domain SSL. All neural encoders are frozen; only the downstream head (RF or MLP) is trained on target-domain labels. Compact TCNet + RF, pretrained on $\leq$23 subjects, surpasses UKB-SSL variants pretrained on ${\sim}$100k subjects on both datasets, despite using orders-of-magnitude less pretraining data.}
  \label{tab:ssl_pretrain}
  \small
  \setlength{\tabcolsep}{3.5pt}
  \begin{adjustbox}{max width=\linewidth}
  \begin{tabular}{llcc|cc|cc}
    \toprule
    \textbf{Method} & \textbf{Pretrain Data} & \textbf{Subject count}& \textbf{Setting} & \multicolumn{2}{c|}{\textbf{UCI-HAR}} & \multicolumn{2}{c}{\textbf{PAMAP2}} \\
     & & & & mF1 & Acc & mF1 & Acc \\
    \midrule
    Raw RF-TSF & N/A & N/A & Scratch & 92.64 & 92.74 & 89.39 & 89.92 \\
    \midrule
    UKB-SSL + RF  & UK Biobank & ${\sim}$100k subjects & Frozen $\to$ RF & 88.96 & 89.04 & 88.04 & 87.84 \\
    UKB-SSL + MLP & UK Biobank & ${\sim}$100k subjects & Finetune & 92.60 & 92.50 & 87.18 & 88.31 \\
    Random Init + MLP & None & N/A & Scratch & 90.43 & 90.40 & 87.90 & 89.35 \\
    \midrule
    \textbf{Compact TCNet + RF (ours)} & UCI-HAR (train) & 23 & Frozen $\to$ RF & \textbf{93.79} & \textbf{93.69} & 89.45 & 89.77 \\
    \textbf{Compact TCNet + RF (ours)} & PAMAP2 (train) & 8 & Frozen $\to$ RF & 90.44 & 90.57 & \textbf{90.23} & \textbf{90.70} \\
    \bottomrule
  \end{tabular}
  \end{adjustbox}
\end{table*}

\textbf{Large-scale latent pretraining can underperform on structured HAR benchmarks.}
Figure~\ref{fig:ssl_pretrain_summary} summarizes Table~\ref{tab:ssl_pretrain} through two comparisons per dataset. First, Compact TCNet + RF improves over Raw RF-TSF on both UCI-HAR (93.79 vs.\ 92.64 mF1) and PAMAP2 (90.23 vs.\ 89.39 mF1), which shows that anchor-guided pretraining adds value beyond static handcrafted features. Second, under frozen-feature evaluation, UKB-SSL + RF falls below Raw RF-TSF on both datasets (UCI-HAR: 88.96; PAMAP2: 88.04), and UKB-SSL + MLP does not surpass the Compact TCNet representation. The frozen UKB embedding therefore does not appear to preserve the TSF-like cues retained by the handcrafted baseline and the anchor-guided encoder.

\begin{figure*}[t]
  \centering
  \includegraphics[width=\linewidth]{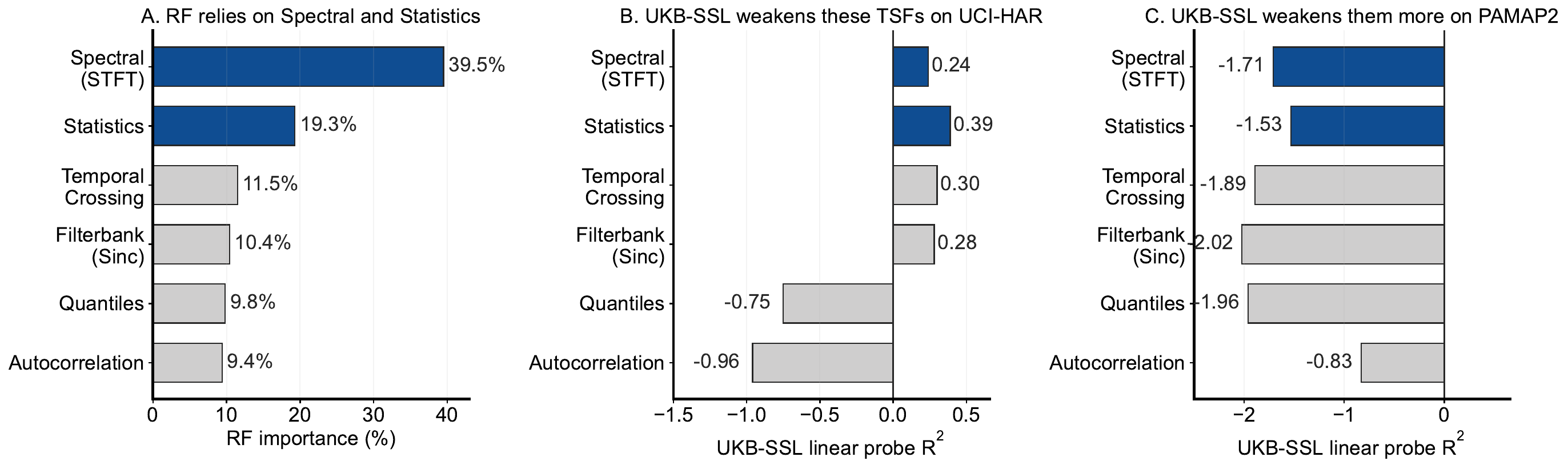}
  \caption{Importance--encodability mismatch for the UKB-SSL encoder. \textbf{Panel A:} RF importance confirms that Spectral and Statistics together account for more than 55\% of discriminative power. \textbf{Panels B \& C:} Linear probe $R^2$ shows that these families are only weakly recoverable from frozen UKB-SSL embeddings on UCI-HAR (Spectral: 0.24; Statistics: 0.39) and are negative on PAMAP2 (Spectral: $-1.71$; Statistics: $-1.53$; all six families negative). The Random Init control retains positive $R^2$ for some families on PAMAP2, indicating that cross-domain pretraining, not architectural bias alone, removes TSF structure under domain mismatch.}
  \label{fig:ukbssl_probe_summary}
\end{figure*}

\textbf{Linear probing explains the gap.}
Table~\ref{tab:probe_ssl} and Fig.~\ref{fig:ukbssl_probe_summary} report how well frozen UKB-SSL representations linearly predict each TSF family using Ridge regression $R^2$. On UCI-HAR, UKB-SSL has mean $R^2 = -0.08$, with Autocorrelation at $-0.96$ and Quantiles at $-0.75$. On PAMAP2, all six families have negative $R^2$ (mean $= -1.66$). This indicates that the frozen embedding does not preserve TSF structure in a linearly accessible form. The random-initialization control retains positive $R^2$ for some families on PAMAP2, which suggests that pretraining under domain mismatch actively shifts representations away from these cues.

\begin{table}[h]
  \caption{Linear probe $R^2$ (ridge regression from frozen encoder to each TSF family) for UKB-SSL and a random-initialization control. On UCI-HAR, UKB-SSL achieves mean $R^2 = -0.08$; on PAMAP2, all six families are negative (mean $= -1.66$), indicating that cross-domain pretraining on single-wrist accelerometry reorganizes representations away from handcrafted feature structure. Negative $R^2$ denotes performance worse than constant-mean prediction.}
  \label{tab:probe_ssl}
  \centering
  \small
  \setlength{\tabcolsep}{3.5pt}
  \begin{adjustbox}{max width=\linewidth}
  \begin{tabular}{lcc|cc}
    \toprule
    & \multicolumn{2}{c|}{\textbf{UCI-HAR}} & \multicolumn{2}{c}{\textbf{PAMAP2}} \\
    \textbf{TSF Family} & UKB-SSL & Rand.Init & UKB-SSL & Rand.Init \\
    \midrule
    Spectral (STFT)     &  0.24 & $-$1.78 & $-$1.71 &  0.19 \\
    Statistics          &  0.39 &  0.84 & $-$1.53 &  0.29 \\
    Temporal Crossing   &  0.30 &  0.49 & $-$1.89 & $-$1.36 \\
    Filterbank (Sinc)   &  0.28 &  0.97 & $-$2.02 &  0.97 \\
    Quantiles           & $-$0.75 & $-$0.85 & $-$1.96 & $-$0.10 \\
    Autocorrelation     & $-$0.96 & $-$0.05 & $-$0.83 &  0.48 \\
    \midrule
    \textbf{Mean}       & $-$0.08 &  0.10 & $-$1.66 &  0.08 \\
    \bottomrule
  \end{tabular}
  \end{adjustbox}
\end{table}

The gap between the two datasets mirrors the domain distance. UKB-SSL has mean $R^2 = -0.08$ on UCI-HAR, a single-device benchmark that is closer to the UKB source domain, but $-1.66$ on PAMAP2, which uses three IMUs and 36 channels. This suggests that TSF structure learned under UKB pretraining transfers poorly to the multi-sensor setting.

\textbf{Architectural inductive bias does not fully explain the gap.}
The Random Init baseline is competitive with UKB-SSL under MLP evaluation and slightly outperforms it on PAMAP2 (87.90 vs.\ 87.18 mF1). Random Init also achieves positive $R^2$ on Filterbank (0.97 on both datasets) and Autocorrelation (0.48 on PAMAP2), which indicates that the ResNet architecture alone exposes some TSF-related structure. Under domain mismatch, pretraining shifts representations away from these cues.

\textbf{Anchor-guided pretraining is data-efficient and task-aligned.}
Compact TCNet is pretrained in-domain on only 23 subjects (7,352 windows) for UCI-HAR or 8 subjects (12,699 windows) for PAMAP2, yet it surpasses both UKB-SSL + RF and UKB-SSL + MLP on the matching target dataset. The pretext tasks detect time reversal, permutation, and warping. These tasks reward encoding statistical regularities such as energy profiles, crossing rates, and autocorrelation structure, which are also captured by handcrafted TSFs and may help explain the data efficiency.

\textbf{Implication for anchor-guided feature learning.}
These results indicate that large-scale latent pretraining is not uniformly superior for structured HAR under domain shift, and that implicit recovery of TSF-like structure can be difficult when the pretraining distribution is misaligned. Compact TCNet shows that a lightweight, task-aligned encoder can surpass a purely handcrafted pipeline in this frozen-feature setting. We treat this as supporting evidence for the broader representation argument rather than as the paper's main contribution: the full TCNet model is still a supervised model that keeps handcrafted TSFs explicit and adaptable during learning.

\section{Discussion}

The main contribution of this paper is not simply a new hybrid HAR architecture. More fundamentally, it proposes a different way to use handcrafted time-series features in wearable recognition systems. In standard pipelines, handcrafted TSFs are typically treated as fixed preprocessing outputs, while deep models learn fully latent representations directly from raw signals. TCNet explores a third option: keeping handcrafted TSFs explicit as \emph{feature anchors} and adapting them with neural context inside the model. The empirical results suggest that this formulation is effective because it preserves a strong statistical scaffold while allowing input-dependent refinement.

This interpretation is supported by three complementary forms of evidence. First, TCNet outperforms strong deep and hybrid baselines on four of the five benchmarks, including the hardest settings in our evaluation. Second, the ablation study shows that the improvement does not come from branch aggregation or additional capacity alone; the largest single drop occurs when anchor correction is removed. Third, the feature-space analyses show that the correction module typically preserves anchor semantics while tightening or rescaling class-relevant structure, and that several discriminative TSF families are not reliably recoverable from standard latent embeddings. Taken together, these results support a representation-level conclusion: for HAR, handcrafted motion statistics can remain useful \emph{inside} the model, provided they are allowed to adapt.

This design also has implications for trustworthy ubiquitous computing. In many wearable applications, especially clinical or safety-critical ones, it is often valuable to inspect intermediate evidence rather than rely only on final predictions~\cite{xai_wearables2024}. TCNet does not solve interpretability in a general sense, but it does offer a more auditable intermediate representation than a purely latent encoder. The model keeps feature families explicit throughout the forward pass and exposes the correction variables that determine how those anchors are modified. This makes it easier to ask not only \emph{what} the model predicted, but also \emph{which kinds of motion statistics were preserved, amplified, or suppressed} in reaching that prediction.

More broadly, the results suggest a lesson beyond wearable HAR. Domain priors are often incorporated either as fixed preprocessing or as weak architectural bias. Our findings point to a more useful middle ground: priors may be most effective when they remain visible as structured intermediate representations and are refined, rather than erased, by learning.

\subsection{Limitations}

TCNet is currently evaluated only on IMU-based HAR with accelerometer, gyroscope, and magnetometer signals. Extending the feature-anchor formulation to other sensing modalities, such as ECG, EMG, or PPG, will require modality-specific feature families and context extractors. In addition, the current model treats the anchor set as fixed. Learning a smaller, task-specific, or sparsity-controlled anchor vocabulary may further improve efficiency and sharpen interpretability.

\section{Conclusion}

This paper addresses a representation gap in wearable HAR. Handcrafted TSFs are explicit and efficient, but in most systems they remain fixed. Deep representations are adaptive, but they are usually latent and difficult to inspect. TCNet bridges this gap by treating handcrafted TSFs as \emph{feature anchors}: explicit statistical representations that remain inside the model and are refined by context-conditioned correction rather than replaced by opaque latent features.

Across five HAR benchmarks, TCNet achieves the strongest overall performance on the more challenging datasets and improves substantially over prior hybrid approaches on USC-HAD, Daphnet, and MHealth. More importantly, the supporting analyses explain where these gains come from. The ablation study shows that anchor correction is the dominant contributor; the raw-versus-corrected feature analysis shows that TCNet typically sharpens class structure without destroying anchor semantics; and the probe results show that several highly useful TSF families are not reliably accessible in standard latent embeddings. Together, these findings support the central claim of the paper: handcrafted motion statistics do not need to be discarded in order to benefit from deep learning. They can instead serve as explicit anchors that a lightweight neural model adapts in feature space.

The broader implication is that structured domain priors can remain valuable in modern learning systems when they are preserved as visible intermediate representations rather than pushed entirely into preprocessing or hidden latent space. For wearable HAR, this yields a model that is compact, accurate, and more auditable. More generally, it suggests a design principle for sensor learning: when a handcrafted representation already captures meaningful structure, the goal of the network may be not to relearn it from scratch, but to correct and refine it.
\begin{acks}
This research is supported by the DUSON-Pratt Pilot Program: Nursing and Engineering Science:
Creating a Transdisciplinary, United Scientific Community; NSF under grants CNS-2112562 and OAC-2503010. 
\end{acks}

\bibliographystyle{ACM-Reference-Format}
\bibliography{sample-base}

\end{document}